\documentclass[fleqn,usenatbib]{mnras}
\usepackage{newtxtext,newtxmath}
\usepackage[T1]{fontenc}
\usepackage[utf8]{inputenc}
\usepackage{natbib}
\usepackage{graphicx}
\usepackage{mathtools}
\usepackage{amsmath}
\usepackage{epsfig}
\usepackage{natbib}
\usepackage{multirow}
\usepackage{rotating}
\usepackage{longtable}
\usepackage{verbatim}
\usepackage{url}
\usepackage{hyperref}
\usepackage{CJK}
\usepackage{caption} 
\usepackage{subcaption} 

\DeclareRobustCommand{\VAN}[3]{#2}
\let\VANthebibliography\thebibliography
\def\thebibliography{\DeclareRobustCommand{\VAN}[3]{##3}\VANthebibliography}

\title[Exoplanet Orbital Distribution I]{Exoplanet Orbital Distribution around FGK Sun-$\odot$-like Host Stars I\\
  \large planet occurrence rate derived from the \emph{Kepler Mission} and theoretical interpretations from planet formation}

\author[L. Zeng et al.]{Li Zeng,$^{1}$\thanks{E-mail: li.zeng@geo.uio.no (PHAB, UiO), astrozeng@gmail.com}
Stephanie C. Werner,$^{1}$
Stein B. Jacobsen,$^{2}$
Elena Mamonova,$^{1}$
Reidar G. Trønnes,$^{3,1}$
Ramon Brasser,$^{4,1}$
\\
$^{1}$Centre for Planetary Habitability (PHAB), University of Oslo, 0315 Oslo, Norway\\
$^{2}$Department of Earth and Planetary Sciences, Harvard University, Cambridge, MA 02138\\
$^{3}$Natural History Museum, University of Oslo, Sars gate 1, 0562 Oslo, Norway\\
$^{4}$Konkoly Observatory, HUN-REN CSFK, MTA Centre of Excellence; Konkoly Thege Miklos St. 15-17, H-1121 Budapest, Hungary
}
\date{Accepted XXX. Received YYY; in original form ZZZ}

\pubyear{\the\year{}}

\begin{document}
\label{firstpage}
\pagerange{\pageref{firstpage}--\pageref{lastpage}}
\maketitle

\begin{abstract}
    Recent astronomical observations, in particular from the \emph{Kepler} and \emph{TESS} missions and their related follow-ups, have revealed an abundance of exoplanets in the size range between Neptune (4 R$_{\oplus}$) and Earth (1 R$_{\oplus}$), as well as a low occurrence rate of planets around twice the radius of Earth (2 R$_{\oplus}$). This paper uses statistical methods, in particular, the \emph{survival function analysis}, to address the known exoplanet population observed mainly from the \emph{Kepler}'s primary mission, in order to mathematically elucidate the orbital distributions (expressed in either the orbital period \emph{P} or the orbital semi-major axis \emph{a}), \emph{for each of the host stars}, both collectively and separately for the planets grouped into various radius bins. We uncover a log-uniform distribution for the majority of planets except for the giants. Based on the results of the statistics, we then visit several possible formation scenarios and pathways for planets in different size ranges, in order to explain the results from a theoretical point-of-view. 
\end{abstract}


\section{Introduction}

The growing census of exoplanets discovered over the past three decades has revealed that planetary systems around FGK Sun-like stars exhibit a striking diversity in both architecture and orbital configuration. Rather than resembling the Solar System, most FGK hosts display a highly non-uniform distribution of planetary orbits, with distinct populations that reflect the underlying physics of planet formation, as well as the biases of current detection techniques. Close in super-Earths and sub-Neptunes with orbital periods shorter than $\sim$100 days are now known to be the most common class of planets, some forming compact multi-planet systems that have no direct analog in our own planetary neighborhood. In contrast, hot-Jupiter gas giants on orbits of only a few days are intrinsically rare, yet their existence provides compelling evidence for significant orbital migration. At larger separations, giant planets tend to cluster near the expected location of the protoplanetary disk's ice line, typically between 1-3 AU, consistent with core accretion theory. Between these regimes lies a relative deficit of giant planets at intermediate periods, often referred to as the "period valley," which likely encodes information about disk dispersal timescales and dynamical evolution.

Among thousands of exoplanets discovered so far, a substantial number plots between the gas giants and rocky exoplanets on a radius histogram, hinting at different structures and compositions for these mid-sized exoplanets. Although the overall radius distribution of Kepler planet data is close to log-normal, it is bi-modal in the smaller planet range with a gap at $\sim$1.8 Earth radii, as first reported by~\citep{Fulton2017ThePlanets} ~ \citep{LPSC2017:Zeng2017PlanetFormation}~\citep{Zeng2017RNAAS} ~ \citep{VanEylen2017AnRocky}~\citep{Zeng2018SurvivalMNRAS} ~ \cite{PNAS:Zeng2019}.

The most common planets appear to be planets with radii approximately equal to 2 to 3 Earth radii, but these planets are not present in our solar system. These planets are called sub-Neptunes. They are naturally bounded on one side by the "radius valley" and on the other side by the "radius cliff"~\citep{Lissauer2023, Dattilo2023, Chakrabarty2026, Aguichine2021, Ho2023}. There may indeed be diversity in composition among the sub-Neptunes. Their variable bulk densities may be indicative of rocky cores with significant gaseous envelopes or water worlds with large fractions of water and steam-dominated envelopes of high mean-molecular-weight or hydrogen-rich envelopes~\citep{Aguichine2025, Benneke2024, Piaulet-Ghorayeb2024, Werlen2026, Werlen2025, Gupta2025, Rogers2010ACompositions, Fernndez-Rodrguez2026, Cooke2026, Madhusudhan2021, Madhusudhan2025, Burn2024, Dorn2026, Luo2024, Lichtenberg2025, Zeng2021}. 



Interpreting this orbital distribution requires disentangling the astrophysical structure from observational selection effects. Transit surveys are most sensitive to short period planets, while radial velocity programs preferentially detect massive planets and require long baselines (long observational periods) to probe wider orbits. As a result, the observed distribution represents a convolution of true planetary distribution with the limitations of current instrumentation. Nevertheless, emerging patterns provide critical constraints on models of planet formation, migration, and long-term dynamical stability. Understanding the orbital architectures of planets around FGK stars is therefore essential not only for reconstructing the processes that shape planetary systems but also for placing the Solar System in its broader galactic context.

Previous efforts have been exerted on the statistics or occurrence rates of exoplanets with respect to the (1) host-stellar types, the (2) planet types, sizes or masses, and the (3) planet orbital semi-major axes or periods, see~\citep{Dattilo2023, Dattilo2024} and references therein. Here, we focus on the planet orbital semi-major axis distribution, with broad binning of planet sizes. 

The orbital semi-major-axis relations indicate the physical separation between the host star and the planet. As we will demonstrate, the orbital distribution function expressed in orbital semi-major axis also shows a simpler form, which points towards various planet formation mechanisms. 

\section{Method}
\emph{Survival Function} (\textbf{SF}), also known as the \emph{complementary Cumulative Distribution Function} (\textbf{cCDF}), is applied to analyze the planet orbital period (\emph{P}) and semi-major axis (\emph{a}) distribution of the \emph{Kepler} sample first. Overall, exoplanets have an approximate \emph{uniform} Probability Density Function (\textbf{PDF}) in the \emph{logarithm} of the orbital semi-major axis ($\ln{(a)}$) or equivalently of the orbital period ($\ln{(P)}$) beyond an inner pivotal point of $\sim$0.05 AU.

Furthermore, we divide the exoplanets into four broad size bins and compare their orbital distribution in each size bin. We find the following. 
\begin{itemize}
    \item 1-2 R$_{\oplus}$: \textbf{PDF} uniform in $\ln{(a)}$, with an inner pivot point at $\sim$0.05 AU.
    \item 2-4 R$_{\oplus}$: \textbf{PDF} uniform in $\ln{(a)}$, with an inner pivot point at $\sim$0.1 AU.
    \item 4-10 R$_{\oplus}$: \textbf{PDF} uniform in $\sqrt{a}$, inside $\sim$0.5 AU.
    \item $>$10 R$_{\oplus}$: \textbf{PDF} uniform in $\sqrt{a}$, inside $\sim$0.5 AU.
\end{itemize}

\section{Survival Function}

\emph{Survival Function} (\textbf{SF}), also known as the \emph{complementary Cumulative Distribution Function} (\textbf{cCDF}), is useful for analyzing the tail of a diminishing distribution~\citep{Clauset2009Power-LawData}~\citep{Feigelson1985StatisticalDistributions}~\citep{Virkar2014Power-lawData}. 

\begin{equation}
    \textbf{SF} \equiv \textbf{cCDF} \equiv 1 - \textbf{CDF} \equiv 1 - \int_{a} \textbf{PDF}(a) \cdot da
\end{equation}

Here we define \textbf{SF} as the cumulative number of exoplanets at orbital distances greater than $a$ (or orbital period greater than $P$) as a function of the variable $a$ (or $P$). We present the result in a log-log plot, that is, $\ln{[SF]}$-vs-$\ln{[a]}$ (or $\ln{P}$) plot.  The data set includes the \emph{Kepler} planet candidates: 4433 in total, from Q1-Q17 of the \emph{NASA Exoplanet Archive}, with false positives excluded~\citep{Akeson2013TheResearch} ~ \citep{Christiansen2025}~\citep{Thompson2017Planetary25}. 

Compared to \textbf{PDF}, \textbf{SF} has the advantage of overcoming large fluctuations in the tail of a distribution due to finite sample sizes~\citep{Clauset2009Power-LawData} ~ \citep{Newman2005PowerLaw}. On such a log-log plot of \textbf{SF} versus the independent variable ($a$), a power-law distribution would appear as a straight line with its slope related to the power-law index, while many other distributions such as normal, log-normal, or exponential distributions would all appear to have sharp pivotal points (upper bounds) in the $a$-axis because those distributions are more bounded within certain limits. Therefore, \textbf{SF} is useful for identifying the \emph{boundaries} that separate different regimes of probabilistic distributions within the dataset. 


\section{Geometric Transit Probability}\label{GeometricTransitProbability}

\begin{figure}
\centering
\includegraphics[width=\columnwidth]{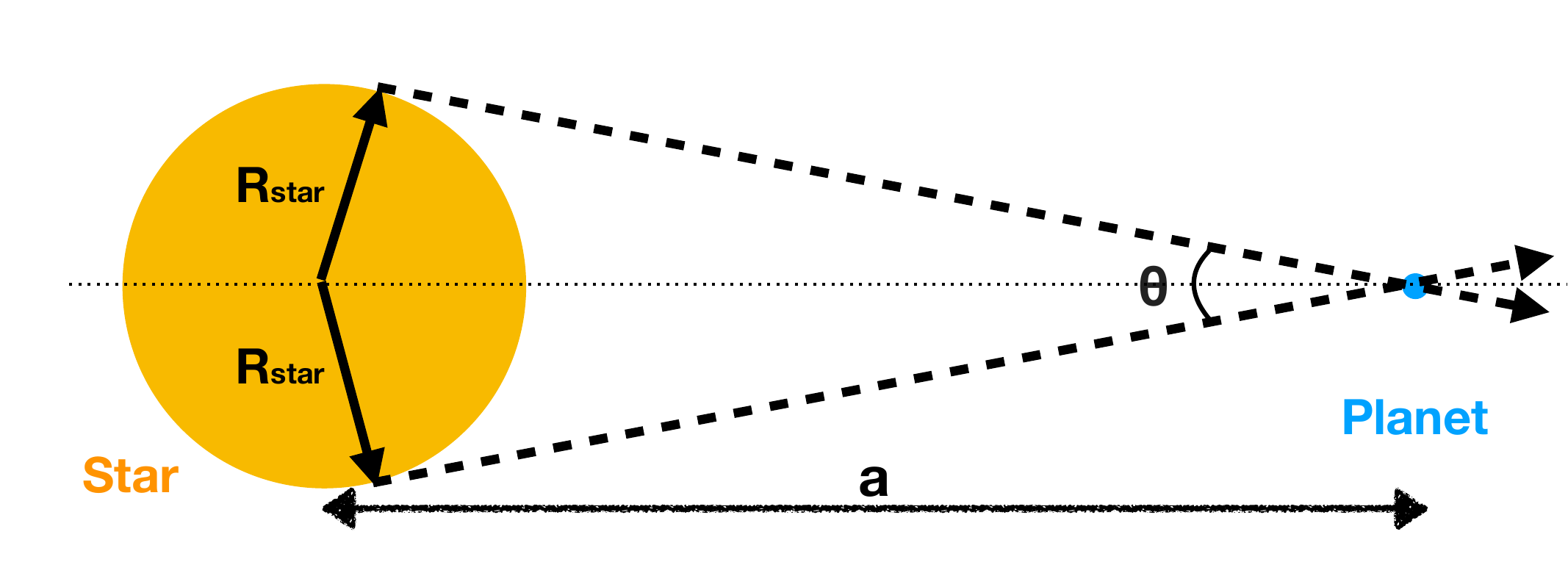}
\caption{Schematic diagram showing the geometry of transit, assuming the radius of the planet (R$_{\text{p}}$) is small compared to the radius of the host star (R$_{\star}$). The orbital period is denoted $P$ in equation~\ref{Eq:TransitProbability}. }
\label{fig:geometric_transit_probability}
\end{figure}

The Geometric Transit Probability is the dominating factor which contributes to the incompleteness and non-detection of exoplanets at greater orbital distances, as surveyed by the transiting method (See Fig.~\ref{fig:geometric_transit_probability}).

Assuming the orientations of orbital planes of planetary systems are random~\citep{Winn2018PlanetSurveys}, then, 
\begin{equation}\label{Eq:TransitProbability}
    \text{Probability of Transit} = \frac{R_{\star}+R_{\text{p}}}{a} \approx \frac{R_{\star}}{a} \approx \frac{R_{\odot}}{a} \propto a^{-1} \sim P^{-2/3}
\end{equation}
because most host stars in the \emph{Kepler} catalog are solar-type FGK main-sequence stars.

Therefore, the Geometric Transit Probability can be corrected for the \textbf{PDF} in \emph{a} or \emph{P} by multiplying by a factor of $(a/R_{\star}) \sim (a/R_{\odot}) \propto a \propto P^{2/3}$. 

\section{Pipeline Incompleteness}\label{PipelineIncompleteness}

The pipeline is the data reduction routine to reduce the raw data collected by the telescope to the positive detection of planet signals. The pipeline incompleteness is another factor contributing to the non-detection of exoplanets~\citep{Burke2017Planet25}~\citep{Christiansen2017Planet25}. Due to the signal-to-noise ratio, the pipeline incompleteness becomes most significant for planets \emph{smaller} than a certain radius threshold and is also correlated with orbital period \emph{P}~\citep{Winn2018PlanetSurveys}. The shorter the orbital period \emph{P} is, the more transits are likely to be observed within a given time frame, and as a result the higher the signal-to-noise ratio is achieved. 


This incompleteness can be probed more rigorously by the injection-recovery test in the pipeline~\citep{Christiansen2020}, which shows a high overall detection efficiency, averaging a 90\%–95\% rate for the end-of-mission version of the Kepler pipeline. This general result indicates a level of incompleteness of a few percent to ten percent for the Threshold Crossing Event (TCE).

Therefore, our conclusion in this manuscript, according to the simple method “survival function” is accurate within about 5-10 percent for most cases. However, we expect to miss some of the smallest planets with size of $\sim$1 to 2 Earth radii, partly because of the low signal-to-noise ratio, and also partly related to the fact that the  "window function", representing the minimum number of transits required for detection, decreases with increasing orbital period. In this study, we do not correct for this, but for the geometric transit probability only. 



As many previous studies provide many details of exoplanet statistics from various parameter-perspective (see, e.g.~\citet{Zhu2021, Dattilo2023, Yang2020, Sayeed2025}), this study focuses on using a simple and clear methodology to produce analytical formulae for the occurrence rate of exoplanets. The analytical formulae can then be used for integration and extrapolation.

\section{Analysis}

\subsection{Survival Function of Orbital Period}\label{SFinP}

\begin{figure}
\centering
\includegraphics[width=\columnwidth]{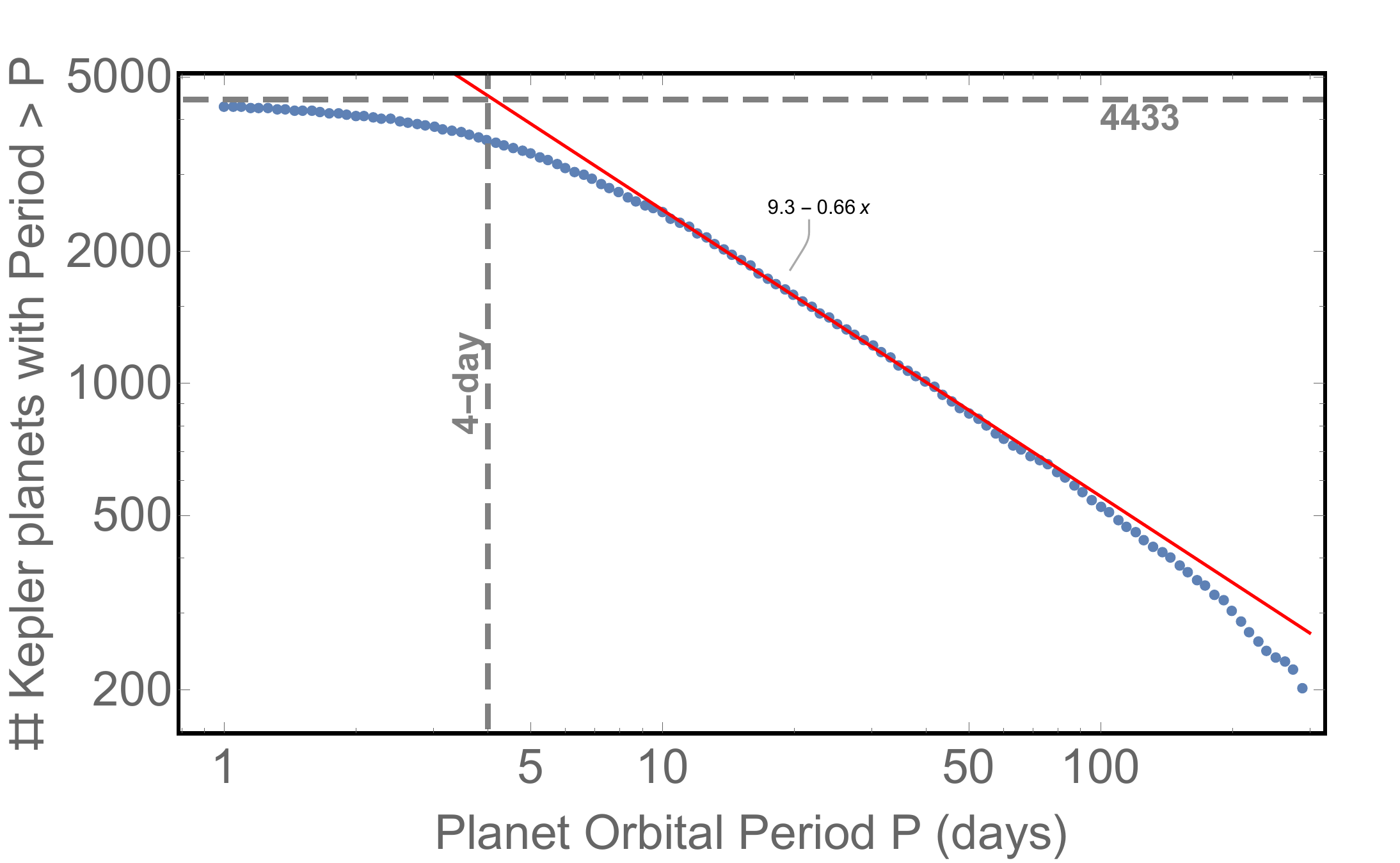}
\caption{Survival Function of Planet Orbital Period Distribution (\emph{P} in days), for Kepler planet candidates, 4433 in total, from the Q1-Q17 NASA Exoplanet Archive~\citep{Akeson2013TheResearch}~\citep{Christiansen2025}~\citep{Thompson2017Planetary25}. X-axis is the orbital period \emph{P}. Y-axis is the number of \emph{Kepler} planet candidates with orbital period larger than a given \emph{P}. The best fit to the data beyond 4-day orbital period is a power-law in logarithmic scale with power index of -2/3 shown as the red line. Here, 4-day can be regarded as the inner pivotal point in \emph{P} of the general assembly of the \emph{Kepler} exoplanets. }
\label{fig:survival_P}
\end{figure}

Fig.~\ref{fig:survival_P} shows the (un-corrected) \textbf{SF} of orbital period \emph{P}, where "un-corrected" stands for "un-corrected of geometric transit probability" as we have discussed in Section~\ref{GeometricTransitProbability}.

\begin{equation}\label{Eq:SF1P}
    \textbf{SF} \propto P^{-2/3} \quad , P>\text{4 days}
\end{equation}

Differentiating \textbf{SF} gives the (un-corrected) \textbf{PDF}: 

\begin{equation}\label{Eq:PDF1P}
    \textbf{PDF} \propto P^{-5/3} \quad , P>\text{4 days}
\end{equation}

Thus, the (un-corrected) number of planets \textbf{dN}$_{\text{obs}}$, detectable by transit from Earth, \emph{per host star}, in the orbital period interval of $[P, P+dP]$ is: 

\begin{equation}\label{Eq:dN1P}
    \textbf{dN}_{\text{obs}} \propto P^{-5/3} \cdot dP \quad , P>\text{4 days}
\end{equation}

Then, according to Eq.~\ref{Eq:TransitProbability}, by multiplying a factor of $P^{2/3}$, the geometric-transit-probability-corrected number of planets \textbf{dN} \emph{per host star} in the orbital period interval of $[P, P+dP]$ is:  

\begin{equation}\label{Eq:dN2P}
    \textbf{dN} \propto P^{-1} \cdot dP = d\ln{P} \quad , P>\text{4 days}
\end{equation}

This is the first indication that the planet population in general is \emph{log-uniformly} distributed in the orbital period \emph{P}, except for an inner pivotal point at $\sim$4-days. The planet population discussed here is for planets larger than 1 R$_{\oplus}$. This lower bound of 1 R$_{\oplus}$ is mainly caused by the pipeline detection sensitivity due to the signal-to-noise ratio as discussed in Section~\ref{PipelineIncompleteness}. 

The next step is to carry out the same analysis in orbital semi-major axis \emph{a}, and we will show that we arrive at the same conclusion of a \emph{log-uniform} distribution of planets in \emph{a}, as a result of \emph{Kepler's Third Law}. Furthermore, we divide the planets into several size bins and carry out the same analysis for each bin, to elucidate the differences between the \emph{Super-Earths} (2 R$_{\oplus} \lesssim$ R$_{\text{p}}\lesssim$ 4 R$_{\oplus}$) and the \emph{Sub-Neptunes} (2 R$_{\oplus} \lesssim$ R$_{\text{p}}\lesssim$ 4 R$_{\oplus}$) and the larger exoplanet population \emph{Super-Neptunes} (R$_{\text{p}}\gtrsim$ 4 R$_{\oplus}$) in terms of their orbital distributions.

\subsection{Survival Function of Semi-major Axis}\label{SFina}

\begin{figure}
\centering
\includegraphics[width=\columnwidth]{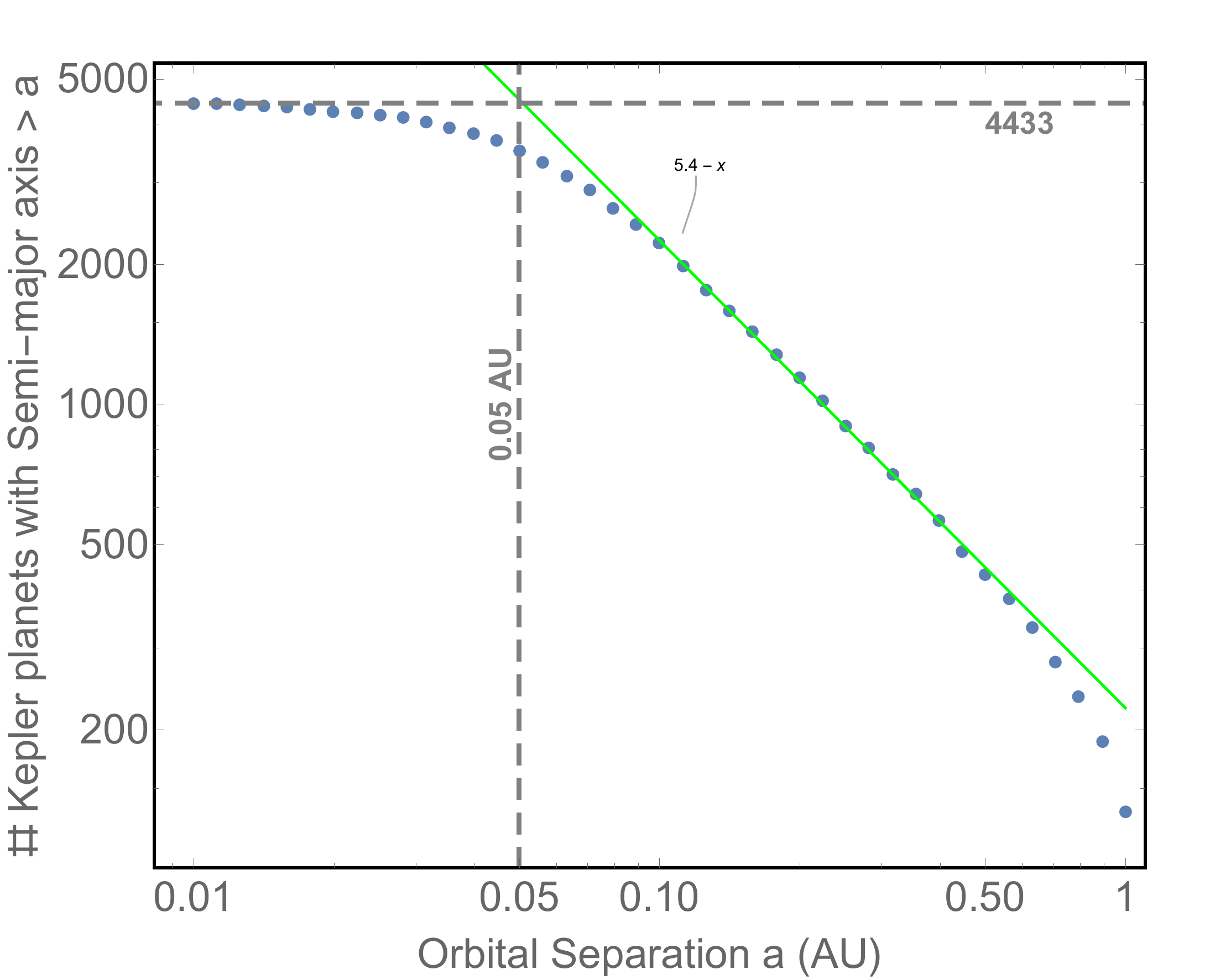}
\caption{Survival Function of Planet Semi-major Axis Distribution (\emph{a} is measured in \emph{astronomical unit} (AU)), for \emph{Kepler} planet candidates, 4433 in total, from the Q1-Q17 NASA Exoplanet Archive~\citep{Akeson2013TheResearch}~\citep{Christiansen2025}~\citep{Thompson2017Planetary25}. X-axis is the semi-major axis \emph{a}. Y-axis is the number of \emph{Kepler} planet candidates with semi-major axis larger than a given \emph{a}. The best fit to the data in the 0.08-0.60 AU range is a power-law in logarithmic scale with power index of -1 shown as the green line. Here, 0.05 AU can be regarded as the inner pivotal point in \emph{a} of the general assembly of the \emph{Kepler} exoplanets. }
\label{fig:survival_a}
\end{figure}

Fig.~\ref{fig:survival_a} shows the (un-corrected) \textbf{SF} of orbital semi-major axis \textbf{a}, where "un-corrected" stands for "un-corrected of geometric transit probability" as we have discussed in Section~\ref{GeometricTransitProbability}.

\begin{equation}\label{Eq:SF1a}
    \textbf{SF} \propto a^{-1} \quad , a>0.05\text{AU}
\end{equation}

Differentiating \textbf{SF} gives the (un-corrected) \textbf{PDF}: 

\begin{equation}\label{Eq:PDF1a}
    \textbf{PDF} \propto a^{-2} \quad , a>0.05\text{AU}
\end{equation}

Thus, the (un-corrected) number of planets \textbf{dN}$_{\text{obs}}$, detectable by transit from Earth, \emph{per host star}, in the infinitesimal semi-major axis interval of $[a, a+da]$ is: 

\begin{equation}\label{Eq:dN1a}
    \textbf{dN}_{\text{obs}} \propto a^{-2} \cdot da \quad , a>0.05\text{AU}
\end{equation}

Then, according to Eq.~\ref{Eq:TransitProbability}, by multiplying a factor of $a^{1}$, the geometric-transit-probability-corrected number of planets \textbf{dN} \emph{per host star}, in the infinitesimal semi-major axis interval of $[a, a+da]$ is: 

\begin{equation}\label{Eq:dN2a}
    \textbf{dN} \propto a^{-1} \cdot da = d\ln{a} \quad , a>0.05\text{AU}
\end{equation}

Therefore, the population of the planet overall is \emph{logarithmically} uniformly distributed on the semi-major axis \emph{a}, except for an inner pivot point at $\sim$0.05 AU. The planet population discussed here is for planets larger than $\sim$1 R$_{\oplus}$. This lower bound of $\sim$1 R$_{\oplus}$ is mainly caused by the sensitivity of pipeline detection due to the signal-to-noise ratio and the "window function" as discussed in Section~\ref{PipelineIncompleteness}. 

According to \emph{Kepler's Third Law} for planets orbiting around a one solar mass (1 M$_{\odot}$) host star, 

\begin{equation}\label{Eq:Kepler3rdLaw}
    a^3 = P^2 \quad , ~\text{\emph{a} in AU and \emph{P} in years}
\end{equation}

or, equivalently, 

\begin{equation}\label{Eq:Kepler3rdLaw2}
    a = (P/365.25)^{2/3} \quad , ~\text{\emph{a} in AU and \emph{P} in days}
\end{equation}

when applied to \emph{P} of 4-day, we have: 
\begin{equation}\label{Eq:Kepler3rdLaw3}
    0.05\text{~AU} = (4/365.25)^{2/3}
\end{equation}

Thus, we have shown that the inner pivot point of $P\sim$ 4 days and $a\sim0.05$ AU are equivalent to each other.
This inner pivotal point of 0.05 AU suggests that "statistically-speaking" planets cannot get closer than this distance to their host stars. 

Because 1 AU = 215 solar radii (R$_{\odot}$), 0.05 AU corresponds to about 11 solar radii (R$_{\odot}$). 

This orbital separation of about $10 \times$ stellar radii away from the host star is generally thought to be where the inner edge of proto-planetary disk truncates, due to various physical mechanisms including magneto-rotational coupling with the proto-star, as well as the termination of planetary gas accretion, caused by the gas disk destruction during the stellar T Tauri stage~\citep{Hartmann2016AnnualReview}. 

\emph{This is a hint that the orbital distribution of these close-in exoplanets may reflect the physical and chemical properties of the disk as well as the disk interaction with the host star}. 

Planets accreted and established within an orbital distance of about ten host star radii would also be strongly irriadiated and eroded, not only by photons but also by ionized plasma in the stellar magnetosphere. In addition to such stellar wind erosion, they would also experience very strong tidal interactions with their host stars:
\begin{equation}\label{Eq:tides}
    \text{Tidal acceleration} \propto \frac{M_{\star}}{a^3}
\end{equation}
Strong tidal interactions will cause rapid orbit circularization and sometimes orbit decay~\citep{Ogilvie2014, Barker2026}.

Therefore, both tides and irradiation may cause relatively short lifetimes and this causes the scarcity of short-period planets ($P\lesssim10$ days) and ultra-short-period planets ($P\lesssim1$ days). 

However, there also seems to exist an outer pivot point of the \emph{logarithmic} uniform distribution in \emph{P} or \emph{a} (Fig.~\ref{fig:survival_P} and Fig.~\ref{fig:survival_a}), which is probably due to the limited observational duration of the \emph{Kepler Space Telescope}, rather than the planet population itself. 

The normal operational mode of the \emph{Kepler Space Telescope} lasted for about three years from December 2009 until the failure of its second reaction wheel in May 2013. During this period, the telescope monitored the same original field of view (FOV), except during the discontinuities
caused by the quarterly rolls of the spacecraft (every 93 days). The data collected during this three-year primary mission are most important for the statistics in this paper. The main data collection during the subsequent K2 mission was mostly outside the original FOV.



As a general criterion, three transits are required in order to confirm a planet candidate. Therefore, $3/3\sim$1 year, or equivalently, $\sim$1 AU orbit, should be regarded as an outer pivotal point, due to the limited observational duration. An outer pivotal point distance and orbital period of about 1 AU and 1 year, respectively, correspond broadly to three transits in the observing period of 3.4 years. This is a physical limit or incompleteness in probing the distant planet population. Most detection of transits beyond this outer pivotal point are single transits, which have less than three repeats and cannot be confirmed further as candidates. In fact, this outer pivotal point can be perceived as the \textbf{SF} starts to fall short of the \emph{log-uniform} distribution beyond the $\sim$100-day orbital period in Fig.~\ref{fig:survival_P} or the 0.5 AU orbital separation in Fig.~\ref{fig:survival_a}. 


Without a limitation in terms of planet detectability at greater orbital distances, this log-uniform distribution in orbital period ($P$) or orbital semi-major axis ($a$) likely continues onward to greater distances from the host stars.

This trend is broadly equivalent to the \emph{Titius-Bode Law}: an empirical (and debated) law of the orbital distribution of solar system planets~\citep{Bode1772, bode, titius, Nieto1972, Nieto1970}, which as a modified power law gives rise to nearly equal spacing of neighboring planets in $\ln{(P)}$ or $\ln{(a)}$.

As a result, PDF \textbf{uniformity} in $\ln{(P)}$ or $\ln{(a)}$ may be the consequence of a \emph{general planet formation process} or a \emph{general planet formation pathway}. This motivates further exploration of the theoretical understanding and modeling of planet formation.

\subsection{Survival Function Binned in Planet Radius}

\begin{figure}
\centering
\includegraphics[width=\columnwidth]{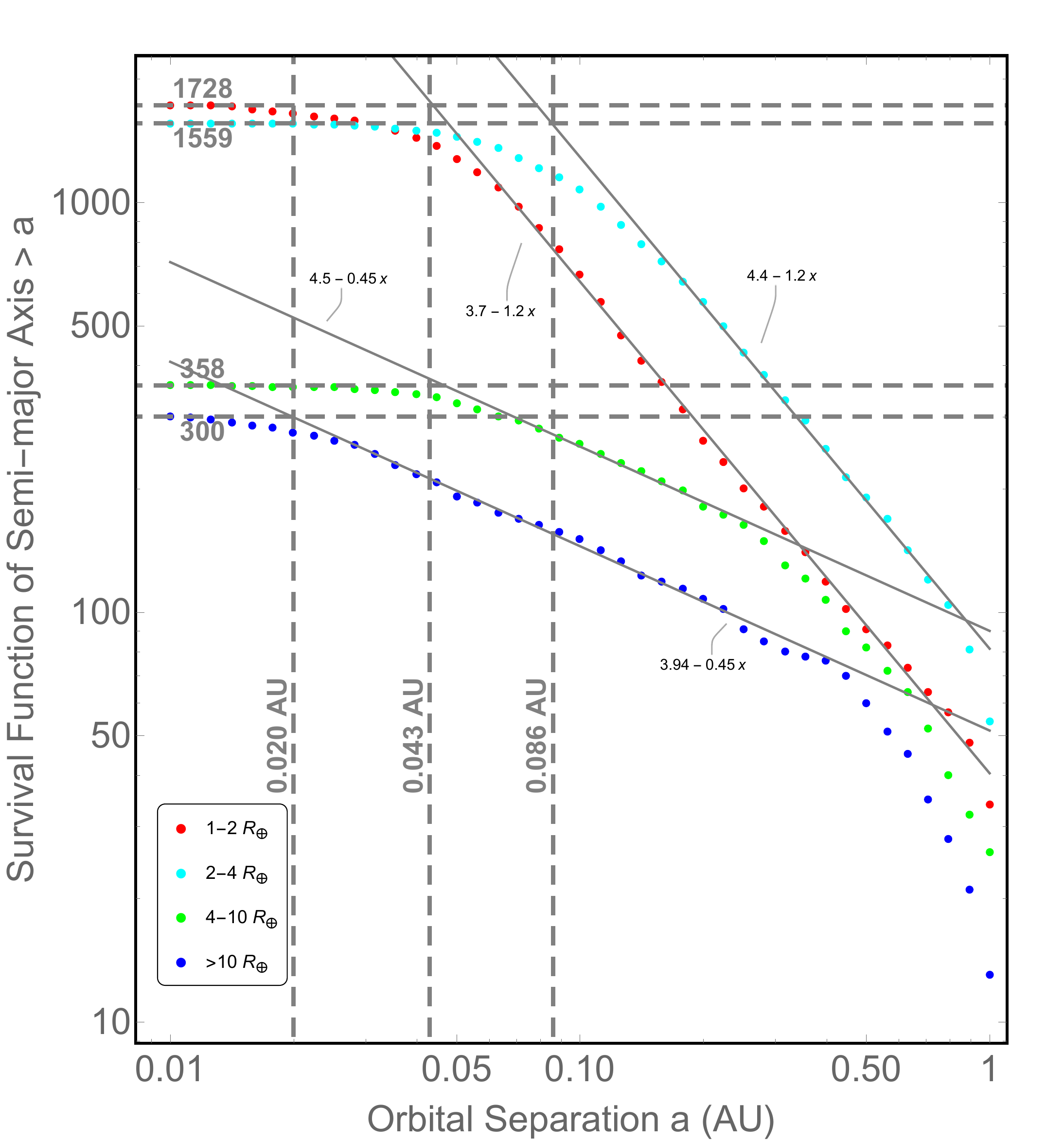}
\caption{Survival Function of Planet Semi-major Axis Distribution (\emph{a} is measured in \emph{astronomical unit} (AU)), for Kepler planet candidates divided into four radius (size) bins: 1-2 R$_{\oplus}$, 2-4 R$_{\oplus}$, 4-10 R$_{\oplus}$, and $>$10 R$_{\oplus}$. X-axis is the semi-major axis \emph{a}. Y-axis is the number of \emph{Kepler} planet candidates with semi-major axis larger than a given \emph{a} in each radius (size) bin. The difference between planets smaller than 4 R$_{\oplus}$ and larger than 4 R$_{\oplus}$ can be perceived.}
\label{fig:survival_a2}
\end{figure}

We divide the \emph{Kepler} planet candidates into four radius (size) bins, according to another independent \textbf{SF} analysis of the planet radius distribution~\citep{Zeng2018SurvivalMNRAS}. The results in Fig.~\ref{fig:survival_a2}, show that within 0.5 AU, the exoplanets smaller than 4 R$_{\oplus}$ have orbital distributions different from those larger than 4 R$_{\oplus}$.


Let us first qualitatively discuss the features of Fig.~\ref{fig:survival_a2}: 

\begin{itemize}
    \item The exoplanet population smaller than 4 R$_{\oplus}$ still has \textbf{PDF} uniform in $\ln{P}$ or $\ln{a}$. In particular, the planet population in the radius bin of (1-2 R$_{\oplus}$) and radius bin of (2-4 R$_{\oplus}$) each show a log-uniform distribution in \emph{a}, manifested by the same negative slope of close to -1 in the log-log plot of \textbf{SF}.
    \item However, these two planet populations show different inner pivotal points. The planet population of (2-4 R$_{\oplus}$) shows an inner pivot point at $\sim$0.1 AU, about twice that of the planet population of (1-2 R$_{\oplus}$) at $\sim$0.05 AU.
    \item Exoplanet populations larger than 4 R$_{\oplus}$ have smaller negative slopes, which means that their \textbf{PDF} is higher at larger orbital distances. In other words, there are more of them ($\gtrsim$4 R$_{\oplus}$), comparatively speaking, at larger orbital distances. 
\end{itemize}


\section{Super-Earths and Sub-Neptunes \\Planets in between one and four Earth radii (1 R$_{\oplus} \lesssim$ R$_{\text{p}}\lesssim$ 4 R$_{\oplus}$)}


The absolute occurrence rate of planets \emph{per host star} in each radius (size) bin can be deduced from Fig.~\ref{fig:survival_a2}. For logarithmic uniform distribution (Eq.~\ref{Eq:dN2a}), the geometric-transit-probability-corrected likelihood \textbf{dN} of finding a planet \emph{per host star} in the infinitesimal semi-major axis interval of $[a, a+da]$ is:  

\begin{equation}\label{Eq:dN3a}
    \textbf{dN} = A \cdot a^{-1} \cdot da = A \cdot d\ln{a} \quad , a_{\text{in}}<a<a_{\text{out}}
\end{equation}

where $a_{\text{in}}$ and $a_{\text{out}}$ are the inner and outer pivot points where this \emph{log-uniform} distribution on the semi-major axis $a$ is valid. Here we introduce a normalization constant $A$ which represents the occurrence rate of planets \emph{per host star}, per unit natural logarithmic interval in the semi-major axis, that is, $e$-fold in \emph{a}. 

The integral $\int$\textbf{dN} gives the \emph{occurrence} of planets within a finite semi-major axis interval, \emph{per host star}. In particular, if we perform the integral from $a_{\text{in}}$ to $a_{\text{out}}$: 

\begin{equation}\label{Eq:dN4a}
    \int \textbf{dN} = \int_{a_{\text{in}}}^{a_{\text{out}}} A \cdot d\ln{a} = A \cdot \ln{\bigg( \frac{a_{\text{out}}}{a_{\text{in}}} \bigg)} 
\end{equation}

However, in order to compare with the detected planet number by transit observation, we need to again multiply a factor of geometric transit probability as in Eq.~\ref{Eq:TransitProbability}. Then 

\begin{equation}\label{Eq:dN5a}
    \textbf{N}_{\text{obs}} = \int \frac{R_{\star}}{a} \cdot \textbf{dN} \approx \int \frac{R_{\odot}}{a} \cdot \textbf{dN} = \int_{a_{\text{in}}}^{a_{\text{out}}} A \cdot R_{\odot} \cdot \frac{da}{a^2} = A \cdot R_{\odot} \cdot \bigg( \frac{1}{a_{\text{in}}} - \frac{1}{a_{\text{out}}} \bigg) 
\end{equation}

Because $a_{\text{out}} \gg a_{\text{in}}$, the total number of planets observed (\textbf{N}$_{\text{obs}}$) by transit \emph{per host star} is: 

\begin{equation}\label{Eq:dN6a}
    \textbf{N}_{\text{obs}} \approx  A \cdot R_{\odot} \cdot \bigg( \frac{1}{a_{\text{in}}} \bigg) = A \cdot \bigg( \frac{R_{\odot}}{a_{\text{in}}} \bigg)
\end{equation}

Eq.~\ref{Eq:dN6a} provides a direct way to extract $A$ from the observed number of transit planets if their orbital distribution is \emph{log-uniform} in the orbital semi-major axis \emph{a} with an inner pivotal point, which is true for planets $\gtrsim$1 R$_{\oplus}$ but $\lesssim$4 R$_{\oplus}$. 

Now, let us put in the actual numbers: 

In the original \emph{Kepler}'s \textbf{FOV}, there are 145,000 intentionally-selected target stars, mostly solar-like FGK main-sequence stars, which were continuously monitored over its normal operational duration of three years. Of them, about four thousand planet candidates with radii $\geqslant$1 R$_{\oplus}$ have been confirmed. 

First of all, according to the analysis in Section~\ref{SFinP} and Section~\ref{SFina}, and combined with Eq.~\ref{Eq:dN6a}, for the overall planet population $\geqslant$1 R$_{\oplus}$, the occurrence rate \emph{per host star} per $e$-fold in \emph{a} is: 

\begin{equation}\label{Eq:dN7a}
    A_{\geqslant 1 \text{R}_{\oplus}} \approx \textbf{N}_{\text{obs}} \cdot \bigg( \frac{a_{\text{in}}}{R_{\odot}} \bigg) \approx \frac{4000}{145,000} \cdot \bigg( \frac{0.05\text{AU}}{1 \text{R}_{\odot}} \bigg) \approx 0.3^{+0.1}_{-0.1} \quad, a_{\text{in}}\approx0.05\text{AU}
\end{equation}

Then separately, for planet population of 1-2 R$_{\oplus}$ and planet population of 2-4 R$_{\oplus}$ each, 

\begin{equation}\label{Eq:dN8a}
    A_{1-2 \text{R}_{\oplus}} \approx \textbf{N}_{\text{obs}} \cdot \bigg( \frac{a_{\text{in}}}{R_{\odot}} \bigg) \approx \frac{1728}{145,000} \cdot \bigg( \frac{0.05\text{AU}}{1 \text{R}_{\odot}} \bigg) \approx 0.1^{+0.03}_{-0.03} \quad, a_{\text{in}}\approx0.05\text{AU}
\end{equation}

\begin{equation}\label{Eq:dN9a}
    A_{2-4 \text{R}_{\oplus}} \approx \textbf{N}_{\text{obs}} \cdot \bigg( \frac{a_{\text{in}}}{R_{\odot}} \bigg) \approx \frac{1559}{145,000} \cdot \bigg( \frac{0.1\text{AU}}{1 \text{R}_{\odot}} \bigg) \approx 0.2^{+0.05}_{-0.05} \quad, a_{\text{in}}\approx0.1\text{AU}
\end{equation}

and, 
\begin{equation}\label{Eq:dN10a}
    0.3 = 0.1 + 0.2
\end{equation}

The conclusion is that although planets of 1-2 R$_{\oplus}$ can get somewhat closer to host star than planets of 2-4 R$_{\oplus}$, however, beyond 0.1 AU, the latter population is about twice as many as the former population, out to at least 0.5 AU, and likely further. Both populations are \emph{log-uniformly} distributed in orbital semi-major axis \emph{a}, and they together comprise the \emph{majority} of the four thousand \emph{Kepler} planet candidates. 

\textbf{WARNING}: this result does not necessarily extend to planets smaller than 1 R$_{\oplus}$ because we do not yet know the \emph{Initial Mass Function} (IMF), or equivalently, \emph{the Initial Radius Function} (IRF), of the small planet population ($\lesssim$1 R$_{\oplus}$) from formation. Thus, there could be many smaller planets compared to larger ones, and any extrapolation downward below 1 R$_{\oplus}$ is inaccurate, due to the detection sensitivity of \emph{Kepler} as discussed in Section~\ref{PipelineIncompleteness}. 

\subsection{Comparison to other works}

\begin{figure}
\centering
\includegraphics[width=\columnwidth]{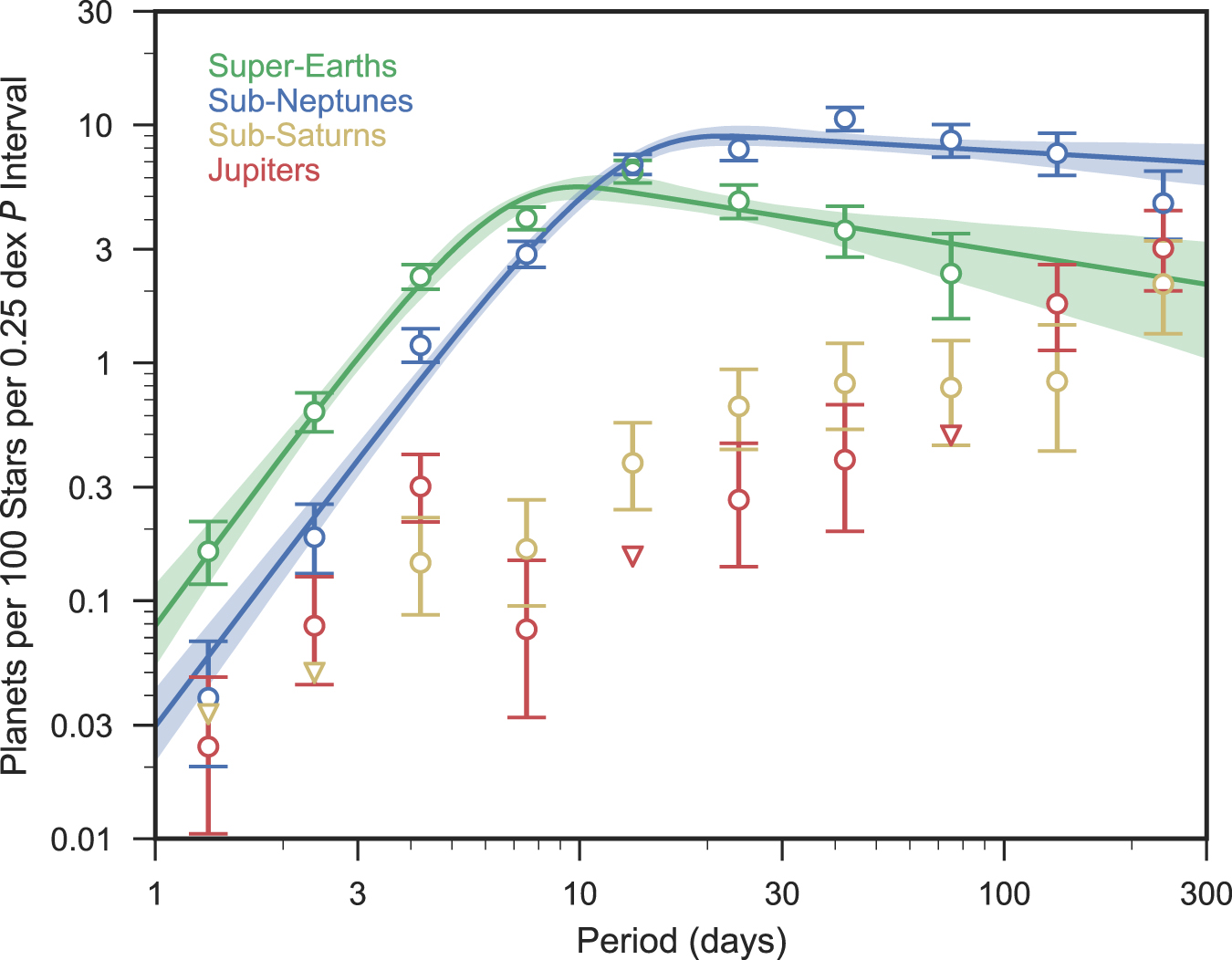}
\caption{~excerpt from~\protect\cite{Petigura2018ThePlanets}. It shows the orbital distribution of exoplanets divided into four radius bins as follows: \emph{Super-Earths} (1-1.7 R$_{\oplus}$), \emph{Sub-Neptunes} (1.7-4 R$_{\oplus}$), \emph{Sub-Saturns} (4-8 R$_{\oplus}$), \emph{Jupiters} (8-24 R$_{\oplus}$), according to their definitions. Geometric Transit Probability and Pipeline Detectability have been corrected by the authors of~\protect\cite{Petigura2018ThePlanets}. }
\label{fig:f1}
\end{figure}

This result is in good agreement with other works (see~\citep{Dattilo2023,Dattilo2024,Sullivan2026,Sayeed2025}), in particular the work by~\cite{Petigura2018ThePlanets}, and also with that of~\cite{Dong2013} and~\cite{Kunimoto2020}. See Fig.~\ref{fig:f1}. The way they divide into bins of planet sizes is slightly different from ours. However, due to the low occurrence rate of planets around 1.7-2 R$_{\oplus}$, known as the \emph{exoplanet radius valley}~\citep{Fulton2017ThePlanets}~\citep{LPSC2017:Zeng2017PlanetFormation}~\citep{Zeng2017RNAAS}~\citep{VanEylen2017AnRocky}~\citep{Zeng2018SurvivalMNRAS}~\cite{PNAS:Zeng2019}, the differences between our statistics are small for the \emph{Super-Earths} (1-2 R$_{\oplus}$) and \emph{Sub-Neptunes} (2-4 R$_{\oplus}$). The natural divide between these two planet groups would be the very bottom of the \emph{exoplanet radius valley}, which may depend on the host stellar type~\citep{Zeng2017RNAAS} ~ \citep{Fulton2018}~\citep{Wu2019}, as well as the physical mechanisms that give rise to it. 

For the clarity of nomenclature, from now on we refer to planets as:
\begin{itemize}
    \item \emph{Super-Earths}: planets larger than Earth but smaller than the \emph{exoplanet radius valley}.
    \item \emph{Sub-Neptunes}: planets larger than the \emph{exoplanet radius valley} but $\lesssim$4 R$_{\oplus}$.
\end{itemize}

Now, let us convert the numbers and units in Fig.~\ref{fig:f1} into the units of $A$ used in our analysis for a proper comparison. There are three major differences in the units that require conversion: 

\begin{itemize}
    \item The Y-axis of Fig.~\ref{fig:f1} is the number of planets \emph{per 100 host stars}. Thus, there is a factor of $100$ when converted to \emph{per host star}. 
    \item The Y-axis of Fig.~\ref{fig:f1} is \emph{per 0.25 dex}, which is $1/4$-interval in log-base-10 ($\lg$). It needs to be converted to a natural-log-base ($\ln$). 
    \item The Y-axis of Fig.~\ref{fig:f1} is per 0.25 dex of \emph{the orbital-period interval}. It needs to be converted to \emph{the semi-major axis interval}, as we decide that it is easier to correct the Geometric Transit Probability from this perspective. The conversion uses \emph{Kepler's Third Law}. 
\end{itemize}

Let us use parameter $(100 \times f)$ to represent the numbers of the Y-axis of Fig.~\ref{fig:f1}, where the factor of $100 \times$ automatically corrects their choice of \emph{per 100 host stars} to our choice of \emph{per host star}: 

\begin{equation}\label{eq1}
f\equiv \frac{d\textbf{N}}{d\lg P}\cdot \frac{1}{4},
\end{equation}

Then 

\begin{eqnarray}\label{eq2}
d\textbf{N} &=& 4\cdot f \cdot \lg P \nonumber\\
   &=& 4\cdot f \cdot \frac{d\ln P}{\ln 10} \nonumber\\
   &=& \frac{2\cdot f}{\ln 10} \cdot 2d\ln P \nonumber\\
   &=& \frac{2\cdot f}{\ln 10} \cdot d\ln{\big(P^2\big)} \nonumber\\
   &\approx& \frac{2\cdot f}{\ln 10} \cdot d\ln{\big(a^3\big)}  \text{Kepler's 3rd Law} \nonumber\\
   &=& \frac{6\cdot f}{\ln 10} \cdot d\ln a \nonumber\\
   &\approx& 2.6 \cdot f \cdot d\ln a \nonumber\\
\end{eqnarray}

Therefore, the conversion from $f$ to $A$ is made by multiplying a factor of $2.6$: 
\begin{equation}\label{eq2b}
    A \approx 2.6 \cdot f
\end{equation}

The flatness of the small planet population ($\lesssim$4 R$_{\oplus}$) in the logarithm of the orbital semi-major axis (or period) is again a prominent feature, as confirmed by Fig.~\ref{fig:f1}. It has an inner pivotal point or turning point at a few$\sim$ten days of orbital period, inside which \textbf{the PDF} attenuates towards the host star [cf.~\cite{Lee2017}].

According to Fig.~\ref{fig:f1}, in the flat part of the distributions, $f_{1-2 \text{R}_{\oplus}} \approx 0.04$ for the \emph{Super-Earth} population, and $f_{2-4 \text{R}_{\oplus}} \approx 0.08$ for the \emph{Sub-Neptune} population. 

Then, the conversion from $f$ to $A$ according to Eq.~\ref{eq2b} yields the following:

\begin{eqnarray*}\label{eq3}
A_{1-2 \text{R}_{\oplus}} \approx 2.6 \times f_{1-2 \text{R}_{\oplus}} = 2.6 \times 4/100 & \approx & 0.1 \nonumber\\
A_{2-4 \text{R}_{\oplus}} \approx 2.6 \times f_{2-4 \text{R}_{\oplus}} = 2.6 \times 8/100 & \approx & 0.2 \nonumber\\
\end{eqnarray*}

Therefore, we have retrieved the result derived earlier in Eq.~\ref{Eq:dN7a}, Eq.~\ref{Eq:dN8a}, and Eq.~\ref{Eq:dN9a}, that is, the average occurrence rate of a particular type of exoplanets, per host star, per logarithmic spacing in the orbital semi-major axis, see below.
\[
\frac{d\textbf{N}}{d\ln a} \approx  
    \begin{cases}
    0.1 & \text{Super-Earths (1-2 R$_{\oplus}$)}, \quad , a_{\text{in}}\approx0.05\text{AU}\\
    0.2 & \text{Sub-Neptunes (2-4 R$_{\oplus}$)}, \quad , a_{\text{in}}\approx0.1\text{AU}.
    \end{cases}
\]



\subsection{Super-Earths and Sub-Neptunes in the Habitable Zone}

The number of \emph{Super-Earths} and \emph{{S}ub-Neptunes} within the habitable zone \emph{per host star} can be estimated from the result above, by assuming that the habitable zone around a typical host star spans a certain range in the orbital semi-major axis (a): 

\[
\textbf{N}_{\text{habitable}} \approx  
    \begin{cases}
    0.1 \times \ln (a_{\text{outer edge}}/a_{\text{inner edge}}) & \text{Super-Earths}, \\
    0.2 \times \ln (a_{\text{outer edge}}/a_{\text{inner edge}}) & \text{Sub-Neptunes}. 
    \end{cases}
\]

where $a_{\text{inner edge}}$ stands for the inner edge of the habitable zone and $a_{\text{outer edge}}$ stands for the outer edge of the habitable zone, measured in \emph{astronomical units} (AU)~\citep{Huang1959, Hart1979, Kasting1993, Kopparapu2013, Kopparapu2014, VanLooveren2025}.

Suppose the log-uniform distribution in $a$ continues out to $\sim1.5$ AU, and we are optimistic that we take $(a_{\text{outer edge}}/a_{\text{inner edge}}) \approx e \approx 2.7 \approx 2-3$, then, the estimate is that \emph{per host star} of FGK-type there is about 0.1 Super-Earth-type planet and 0.2 Sub-Neptune-type planet on average within its habitable zone. 

However, please keep in mind that there could be many more smaller planets ($\lesssim$1 R$_{\oplus}$) within the habitable zone of such a star. If we take the solar system as an example, and only count the planets larger than two Earth radii, then we would have completely missed the inner planets of Venus, Earth, and Mars which lie within the habitable zone, but only count the distant gas giant planets and icy giants. So we need to have a better understanding of the \emph{Initial Mass Function} (IMF) of the planets formed in proto-planetary disks in general, and also a better understanding of the general formation mechanisms of various types of planets. c.f.~\citep{Fernandes2025}

\subsection{Formation Scenario: in-situ or disk migration?}

\subsubsection{in-situ formation}
The flatness of distributions in $\ln{a}$ suggests that for both populations of Super-Earths and Sub-Neptunes that we consider here, their presence and orbital locations are \emph{not} strongly influenced by the host star. This presents a challenge to the photo-evaporation hypothesis (see also~\citep{Loyd2020}). In this hypothesis, the distinction between these two types of planets (Super-Earths versus Sub-Neptunes) shall heavily depend on the level of host-stellar radiation, and thus shall heavily depend on the orbital semi-major axis to the host star.

Here, the question we want to address is what has caused this flat distribution. Could the answer be \emph{in-situ} formation or could the answer be disk migration?

\emph{In-situ} formation is a plausible answer. Changing the orbital semi-major axis of a planet to its host star (once it is formed) requires an immense amount of energy:

\begin{equation}\label{eqOrbitalEnergy}
    \Delta E_{\text{orbit}} = \Delta \left[ -\frac{G M_{\star} \cdot M_{\text{planet}}}{a} \right] = G M_{\star} \cdot M_{\text{planet}} \cdot \Delta \left[ -\frac{1}{a} \right]
\end{equation}

the change in the specific orbital energy ($\epsilon_{\text{orbit}}$, that is, the orbital energy per unit of planet mass) is: 

\begin{equation}\label{eqOrbitalEnergy2}
    \Delta \epsilon_{\text{orbit}} = \Delta \left[ \frac{E_{\text{orbit}}}{M_{\text{planet}}} \right] = G M_{\star} \cdot \Delta \left[ -\frac{1}{a} \right]
\end{equation}

It also requires a great change in the orbital angular momentum ($L_{\text{orbit}}$). c.f.~\cite{Chiang2013, Hansen2012, Hansen2013, Lee2017}: 

\begin{eqnarray}\label{eqOrbitalAngularMomentum}
    \Delta L_{\text{orbit}} &=& \Delta \left[ M_{\text{planet}} \cdot \sqrt{ G \cdot M_{\star} \cdot a \cdot (1-e^2) } \right] \nonumber\\
    &=& \left( M_{\text{planet}} \cdot \sqrt{G M_{\star}} \right) \cdot \Delta \left[ \sqrt{a \cdot (1-e^2) } \right] \nonumber\\
\end{eqnarray}

the change in the specific orbital angular momentum ($l_{\text{orbit}}$, that is, the orbital angular momentum per unit of planet mass) is: 

\begin{equation}\label{eqOrbitalAngularMomentum2}
    \Delta l_{\text{orbit}} = \Delta \left[ \frac{L_{\text{orbit}}}{M_{\text{planet}}} \right] = \left( \sqrt{G M_{\star}} \right) \cdot \Delta \left[ \sqrt{a \cdot (1-e^2) } \right]
\end{equation}

The flatness in the logarithmic of semi-major-axis distribution points to the possibility of in-situ formation of a small planet (with radius less than about four Earth radii) within the proto-planetary disk, as such a planet can form anywhere in the disk. 



\subsubsection{disk migration}
\begin{figure}
\centering
\includegraphics[width=\columnwidth]{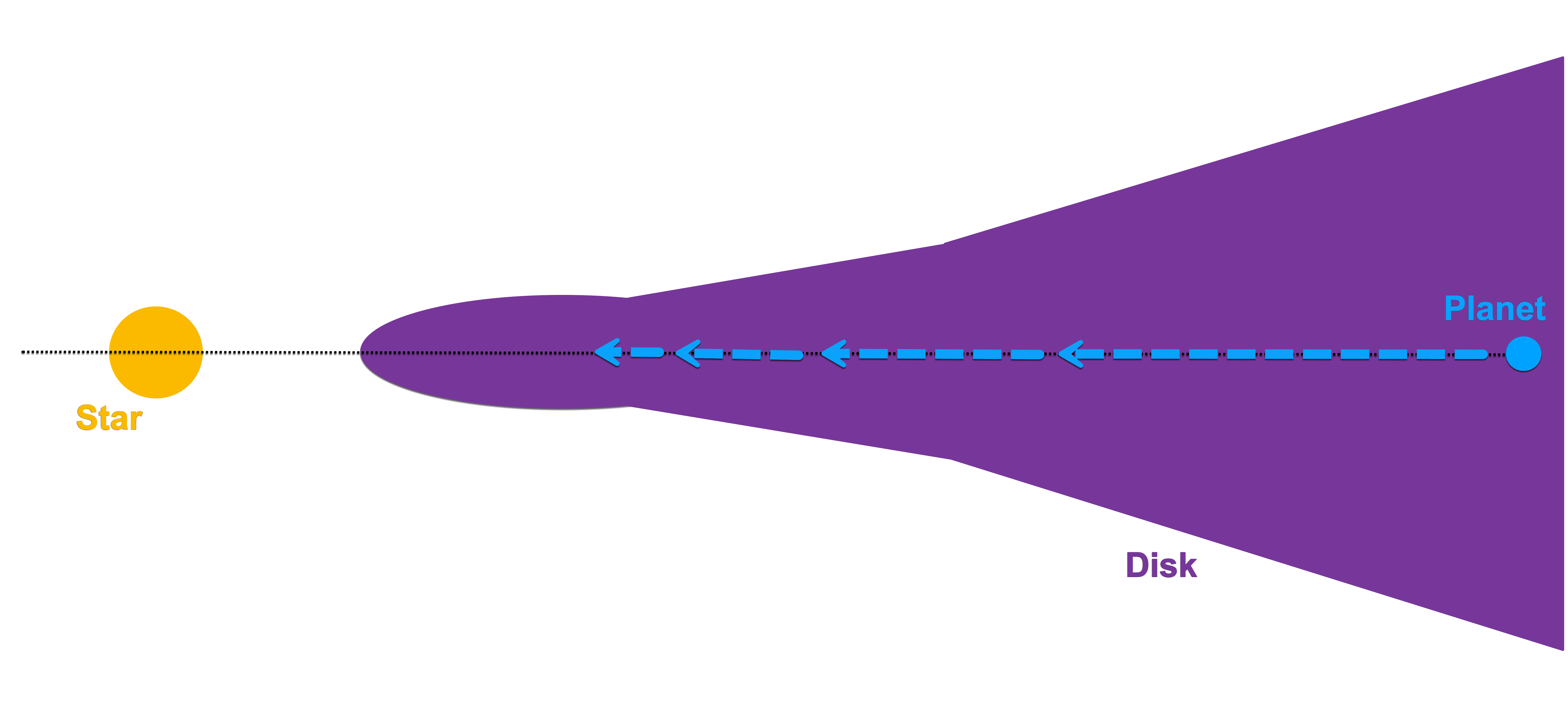}
\caption{Cartoon illustrating a planet migrating towards its host star in a proto-planetary disk, before the disk is completely dissipated. }
\label{fig:f2}
\end{figure}

Compared to the life-span of a star, the proto-planetary disk is a short-lived transient phenomenon. 
So, alternatively, planet could perhaps \emph{migrate} towards the host star when the disk is present. Fig.~\ref{fig:f2} is a heuristic illustration of this idea. Planet migration requires the exchange of orbital angular momentum ($L_{\text{orbit}}$) with disk matter or magnetic fields. Excessive angular momentum could perhaps be transported and then extracted from the inner truncation region near the host star by magneto-rotational coupling through helical bi-polar jets (outflows). These helical bi-polar jets could carry away immense amounts of angular momentum from the inner edge of the disk. These jets shoot in the polar directions perpendicular to the disk-plane, and they have been confirmed by \emph{ALMA} Observations~\citep{Hartmann2016AnnualReview} ~ \citep{Lee2018}~\citep{Lee2020} ~ \citep{Pudritz2019}. 

For simplicity, let us assume that the characteristic (inward) \emph{migration timescale} of a single planet in the disk is constant $\tau_{\text{migration}}$, which is \emph{independent} of its location in the disk and also \emph{independent} of the evolutionary stage of the disk, as long as the disk exists. This gives an exponential decay of the semi-major axis \emph{a} of this planet. The time-dependence of \emph{a} resembles that of a radioactive decay. It is expressed by the following simple differential equation:

\begin{equation}\label{eq2.1}
\frac{1}{a} \cdot \frac{da}{dt} = \frac{\dot a}{a} = \frac{1}{\tau_{\text{migration}}}
\end{equation}

The solution of this equation~\ref{eq2.1} is an exponential decay function in time: 

\begin{equation}\label{eq2.2}
a(t) = a(0) \times \exp{\bigg(-t/\tau_{\text{migration}} \bigg)}
\end{equation}

or, equivalently,

\begin{equation}\label{eq2.3}
\ln{\bigg( \frac{a(t)}{a(0)} \bigg)} =  - \frac{t}{\tau_{\text{migration}}}
\end{equation}

The final orbital semi-major axis ($a_{\text{final}}$) of this planet is where the planet stops migrating, which depends on the initial orbital separation $a(0)$ and the disk lifetime: $\tau_{\text{disk}}$. 

More specifically, it depends on the ratio $\tau_{\text{disk}}/\tau_{\text{migration}}$ . 

The lifespan of the disk $\tau_{disk}$ has a certain distribution. This distribution can be probed by observing the fraction of young stars that have disks from the same young star cluster, which were presumably formed at about the same age. Then, this fraction can be plotted versus the age of the cluster for several clusters (Fig.~\ref{fig:f2b}): 

\begin{figure}
\centering
\includegraphics[width=\columnwidth]{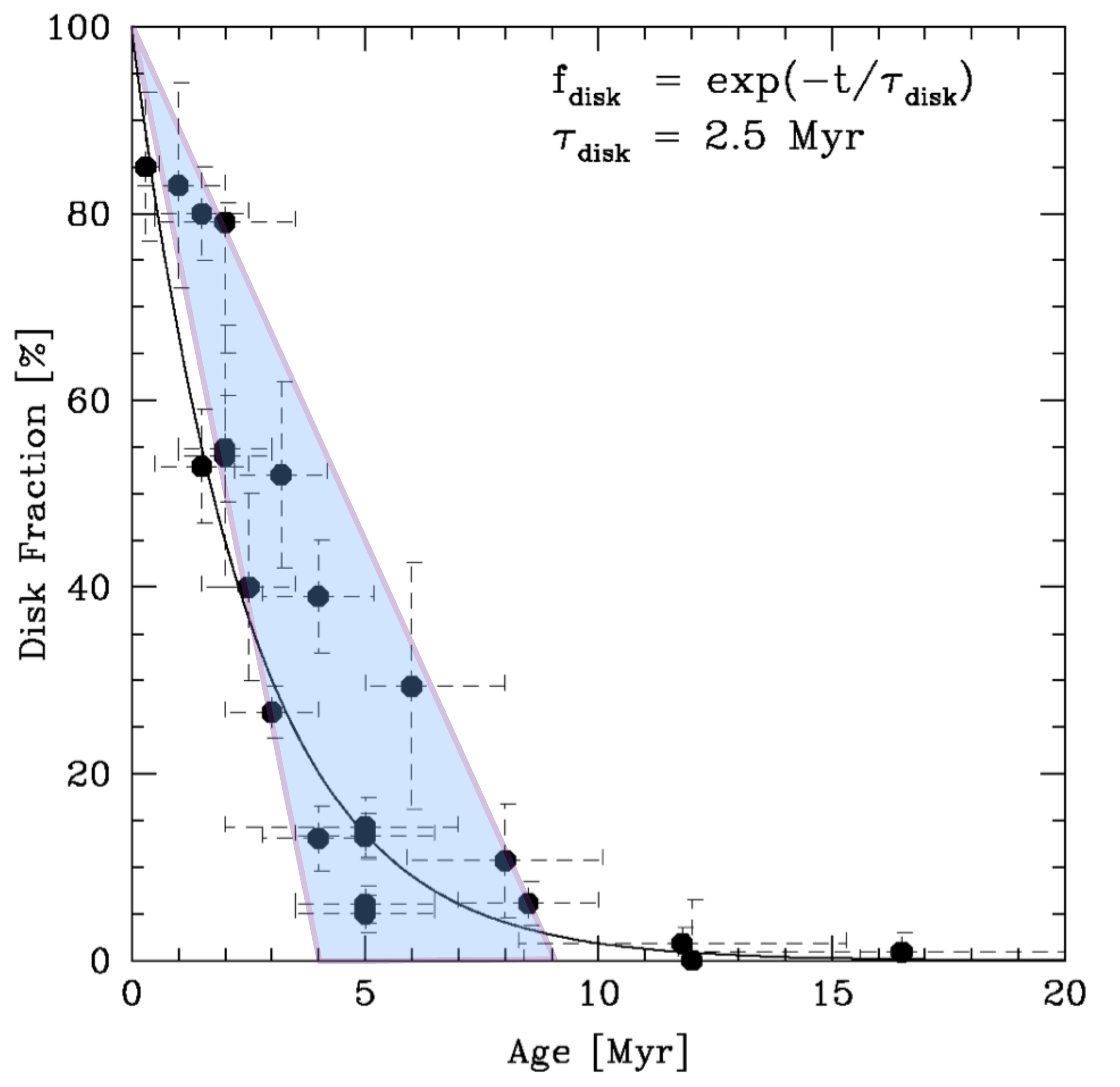}
\caption{~excerpt from~\protect\cite{Mamajek2009InitialDisks}. It shows the age of clusters versus the fraction of stars with primordial disks, either through H$\alpha$ emission or infrared excess diagnostics. Although the authors fit the distribution to an exponential decay curve with $\tau_{\text{disk}} \approx 2.5$ Myr, a linear decay (a straight line) fits equally well with the data within errors. In fact, this disk fraction (percentage) versus age (Myr) plot shows the \emph{cumulative distribution function} \textbf{CDF} (which is linear) of the disk lifespan distribution \textbf{PDF} (which is flat).}
\label{fig:f2b}
\end{figure}

If we take a linear (straight-line) fit of Fig.~\ref{fig:f2b}, then the \textbf{PDF} distribution of disk life $\tau_{\text{disk}}$ is \emph{flat} from 0 up to $5\sim10$ Myr. 
Furthermore, 
\begin{itemize}
    \item if $\tau_{migration}$ is a constant as assumed previously, and it is on the order of 1 Myr (the current type I migration models over-estimate the migration rate by almost two-orders-of-magnitude, as pointed out by~\cite{Kley2012Planet-DiskEvolution}), and
    \item if a \emph{Super-Earth} or \emph{Sub-Neptune} initially forms at around a few AU ($a(0)\sim$ a few AU), and starts inward migration from there during the disk lifespan ($\tau_{\text{disk}}$), and stops migration immediately after the disk is gone, 
\end{itemize}

then, this \emph{flat} \textbf{PDF} in disk lifespan $\tau_{\text{disk}}$ between 0 and $5\sim10$ Myr translates directly into a \emph{log-uniform} distribution of planets on the semi-major axis $a$, due to Eq.~\ref{eq2.3}.

\subsection{Short-Period super-Earths and sub-Neptunes}
Fig~\ref{fig:f1} shows another important feature: the occurrence rate \textbf{PDF} of Short-Period ($P\lesssim10$ days) \emph{Super-Earths} and \emph{Sub-Neptunes}, inside their inner pivotal points of their \emph{log-uniform} distributions, resemble one another in its functional form, and generally agrees with a \emph{power-law} with the same \emph{power-index} of $2.3\approx \frac{7}{3}$ in orbital period $P$, except for a slight parallel shift between these two populations. In terms of parameter $f$ as we have defined earlier: 

\begin{equation}\label{eq3.1}
f \propto P^{7/3} \propto a^{7/2},
\end{equation}

Then, the question arises: What physical mechanism causes this \emph{power-law} of Short-Period \emph{Super-Earths} and \emph{Sub-Neptunes}?

One candidate mechanism is \emph{orbital decay due to tidal interactions}. Perhaps we should also consider the ages of these systems, as tidal interactions may be negligible (or absent) in young systems. 

Planet tides, which are raised by the host star, are mostly responsible for \emph{eccentricity damping} and orbital circularization; while stellar tides, which are raised by the planet, are responsible for the decay of orbital semi-major axis \emph{a}. See, for example,~\cite{Ferraz-Mello2007TidalRe-visited} and~\cite{DobbsDixon2004SpinOrbitPlanets}). The assumption of tidal quality factor $\bf{Q}$ of the host star becomes important as it varies as the planet approaches the host star and becomes frequency-dependent. $\bf{Q}$ of some host stars have been measured from some Hot-Jupiters~\cite{Cameron2018HierarchicalJupiters} and confirm the dependence predicted by the inertial wave dissipation model~\cite{Ogilvie2007TidalStars}. 

Accordingly, the characteristic decay timescale $\tau_{a}$ of the orbit at the semi-major axis \emph{a} due to tides on the host star raised by the planet is:

\begin{equation}\label{eq3.2}
\tau_a \propto a^{13/2} \cdot \frac{1}{M_{\text{planet}}},
\end{equation}

so that 

\begin{equation}\label{eq3.3}
\tau_a \propto a^{7/2} \cdot \left( \frac{a^3}{M_{\text{planet}}} \right) \sim a^{7/2} \cdot \left( \frac{(a/R_{\text{planet}})^3}{M_{\text{planet}}/R_{\text{planet}}^3} \right) \sim a^{7/2} \cdot \left( \frac{(a/R_{\text{planet}})^3}{\rho_{\text{planet}}} \right),
\end{equation}

The quantity in parentheses above has the unit of inverse density. 

\section{Super-Neptunes:\\planets larger than about four Earth radii  (R$_{\text{p}}\gtrsim$ 4 R$_{\oplus}$)}

Fig.~\ref{fig:survival_a2} shows that for \emph{Super-Neptune} planets (4-10 R$_{\oplus}$) and ($>$10 R$_{\oplus}$), their \textbf{SF} slope is close to $-1/2$ inside $\sim$0.5 AU. Therefore, a different normalization scheme is needed to deduce their occurrence rates. Here we introduce the normalization constant $B$, which has the dimension of $1/\sqrt{\text{AU}}$. So, within the applicable range of some $a_{\text{in}}$ and $a_{\text{out}}$, 

\begin{equation}\label{Eq:dN3b}
    \textbf{dN} = B \times d(\sqrt{a}) = B \cdot \frac{da}{2\cdot \sqrt{a}} \quad , a_{\text{in}}<a<a_{\text{out}}
\end{equation}

The integral $\int$\textbf{dN} gives the \emph{occurrence rate} of such planets within a finite semi-major axis interval \emph{per host star}. In particular, if we perform the integral from $a_{\text{in}}$ to $a_{\text{out}}$: 

\begin{equation}\label{Eq:dN4b}
    \int \textbf{dN} = \int_{a_{\text{in}}}^{a_{\text{out}}} B \cdot d(\sqrt{a}) = B \cdot \big( \sqrt{a_{\text{out}}}-\sqrt{a_{\text{in}}} \big)
\end{equation}

and after correcting for the geometric transit probability according to Eq.~\ref{Eq:TransitProbability}, the number of planets observed (\textbf{N}$_{\text{obs}}$) by transit \emph{per host star} is: 

\begin{equation}\label{Eq:dN5b}
    \textbf{N}_{\text{obs}} = \int \frac{R_{\star}}{a} \cdot \textbf{dN} \approx \int \frac{R_{\odot}}{a} \cdot \textbf{dN} = \int_{a_{\text{in}}}^{a_{\text{out}}} B \cdot R_{\odot} \cdot \frac{d\sqrt{a}}{a} = B \cdot R_{\odot} \cdot \bigg( \frac{1}{\sqrt{a_{\text{in}}}} - \frac{1}{\sqrt{a_{\text{out}}}} \bigg) 
\end{equation}

Eq.~\ref{Eq:dN5b} provides a direct way to extract $B$ from the observed number of transit planets if their orbital distribution is \emph{uniform} in $\sqrt{a}$ within a certain range, which is true for planets $\gtrsim$4 R$_{\oplus}$. 

Now, let us put in the actual numbers: 

Again, in \emph{Kepler}'s original \textbf{FOV}, there were about 145,000 solar-like FGK target stars that were continuously monitored for about three years. According to Fig.~\ref{fig:survival_a2},

For the planet population of 4-10 R$_{\oplus}$: $a_{\text{in}} \approx 0.04$ AU and $a_{\text{out}} \approx$ 0.5 AU, 

\begin{equation}\label{Eq:dN8b}
    B_{4-10 \text{R}_{\oplus}} \approx \Bigg( \frac{1\text{AU}/R_{\odot}}{\frac{1}{\sqrt{0.04}} - \frac{1}{\sqrt{0.5}}} \Bigg) \cdot \bigg( \frac{358-120}{145,000} \bigg) \approx 0.1 \times \bigg( \frac{1}{\sqrt{\text{AU}}} \bigg)
\end{equation}

For the planet population of $>$10 R$_{\oplus}$: $a_{\text{in}} \approx 0.02$ AU and $a_{\text{out}} \approx$ 0.5 AU, 
\begin{equation}\label{Eq:dN9b}
    B_{>10 \text{R}_{\oplus}} \approx \Bigg( \frac{1\text{AU}/R_{\odot}}{\frac{1}{\sqrt{0.02}} - \frac{1}{\sqrt{0.5}}} \Bigg) \cdot \bigg( \frac{300-80}{145,000} \bigg) \approx 0.05 \times \bigg( \frac{1}{\sqrt{\text{AU}}} \bigg)
\end{equation}

accordingly, we can estimate the total number of such planets inside 0.5 AU for each case: 

\begin{equation}\label{Eq:dN10b}
   \textbf{N}_{4-10 \text{R}_{\oplus}} \approx B_{4-10 \text{R}_{\oplus}} \times \big(\sqrt{0.5\text{AU}}-\sqrt{0.04\text{AU}} \big) \approx 0.05 \quad, a<0.5\text{AU}
\end{equation}

\begin{equation}\label{Eq:dN11b}
   \textbf{N}_{>10 \text{R}_{\oplus}} \approx B_{>10 \text{R}_{\oplus}} \times \big(\sqrt{0.5\text{AU}}-\sqrt{0.02\text{AU}} \big) \approx 0.03 \quad, a<0.5\text{AU}
\end{equation}

and, 
\begin{equation}\label{Eq:dN12b}
    0.05 + 0.03 = 0.08
\end{equation}

Accordingly, there are only about 0.08 planets of ($>$ 4 R$_{\oplus}$) \emph{per host star} within 0.5 AU. 
In comparison, there are about 0.6 planets (1 R$_{\oplus} <$ R$_{\text{p}}<$ 4 R$_{\oplus}$) \emph{per host star} within 0.5 AU, because  
\begin{align*}\label{Eq:dN12a}
    \textbf{N}_{1-4 \text{R}_{\oplus}} &\approx A_{1-2 \text{R}_{\oplus}} \cdot \ln{\bigg(\frac{a_{\text{out}}}{a_{\text{in}}} \bigg)} + A_{2-4 \text{R}_{\oplus}} \cdot \ln{\bigg(\frac{a_{\text{out}}}{a_{\text{in}}} \bigg)},\\ &\approx 0.1 \times \ln{\bigg( \frac{0.5\text{AU}}{0.05\text{AU}} \bigg)} + 0.2 \times \ln{\bigg( \frac{0.5\text{AU}}{0.1\text{AU}} \bigg)},\\ &\approx 0.6 \quad , a<0.5\text{AU}
\end{align*}

Therefore, the number of planets of ($>$ 4 R$_{\oplus}$) is about an order-of-magnitude lower compared to the planets of (1 R$_{\oplus} <$ R$_{\text{p}}<$ 4 R$_{\oplus}$), inside 0.5 AU. This is an indication that larger planets may have originated at orbital separations from the host star larger than smaller planets. 

\subsection{Comparison to other works}
This result is still in good agreement with other works, in particular, the work by Erik Petigura et al.~\cite{Petigura2018ThePlanets}. See Fig.~\ref{fig:f1}. 

However, in this way they divide up the \emph{Super-Neptunes}: 
\begin{itemize}
    \item \emph{Sub-Saturns}: (4-8 R$_{\oplus}$)
    \item \emph{Jupiters}: (8-24 R$_{\oplus}$)
\end{itemize}

differs somewhat from ours~\citep{Zeng2018SurvivalMNRAS}: 
\begin{itemize}
    \item \emph{Transitional Planets}: (4-10 R$_{\oplus}$)
    \item \emph{Gas Giants}: ($>$10 R$_{\oplus}$)
\end{itemize}

The main results agree with each other. 

Fig.~\ref{fig:f1} shows a notable feature of the \emph{Super-Neptunes}: a gradual and linear increase in the logarithm of the occurrence rate $\ln{(f)}$ over $\ln{(P)}$. Examination of its slope of Sub-Saturns and Jupiters in Fig.~\ref{fig:f1} gives the following approximate dependence: 

\begin{equation}\label{eq4}
    f \propto P^{1/3},
\end{equation}

Then, according to \emph{Kepler's Third Law} once again, it translates to a square-root dependence on the semi-major axis $\sqrt{a}$:

\begin{equation}\label{eq5}
    f \propto a^{1/2},
\end{equation}

Recall the definition of $f$ and its differential relation with the \emph{cumulative number} of planets $\textbf{N}$ \emph{per host star}, according to Eq.~\ref{eq2} and Eq.~\ref{eq2b}, 

\begin{equation}\label{eq6}
    f = \frac{1}{2.6} \cdot d\textbf{N} \cdot \frac{a}{da},
\end{equation}

then, combining Eq.~\ref{eq5} and Eq.~\ref{eq6}, we have
\begin{equation}\label{eq7}
    d\textbf{N} \propto d(\sqrt{a}),
\end{equation}

or equivalently, 
\begin{equation}\label{eq8}
    \textbf{N} \propto \sqrt{a},
\end{equation}

where $\textbf{N}$ represents the \emph{cumulative number} of planets \emph{per host star} from very close to the host star (we take $a_{\text{in}}$ as very small and neglect it here), out to a given semi-major axis $\bf{a}$. Thus, we have retrieved the \emph{uniform-in-}$\sqrt{a}$ distribution of the \emph{Super-Neptunes} from the previous analysis. 

This $\sqrt{a}$-functional dependence is likely a consequence of \emph{the planet-planet scattering} process, which transports a fraction of these \emph{Super-Neptunes}, which may have originated from a few AU or even farther out, to their current orbital separations. We will discuss this idea in the next section. c.f.~\cite{Carrera2019, Adams2003, Chatterjee2008, Raymond2009, Raymond2010, Raymond2020, Rasio1996, Weidenschilling1996, Lin1997, Moorhead2005, Ford2008, Juri2008, Zare1977, Gladman1993, Marchal1982, Chambers1996}

\subsection{Formation Scenario: Planet-Planet Scattering?}

\begin{figure}
\centering
\includegraphics[width=\columnwidth]{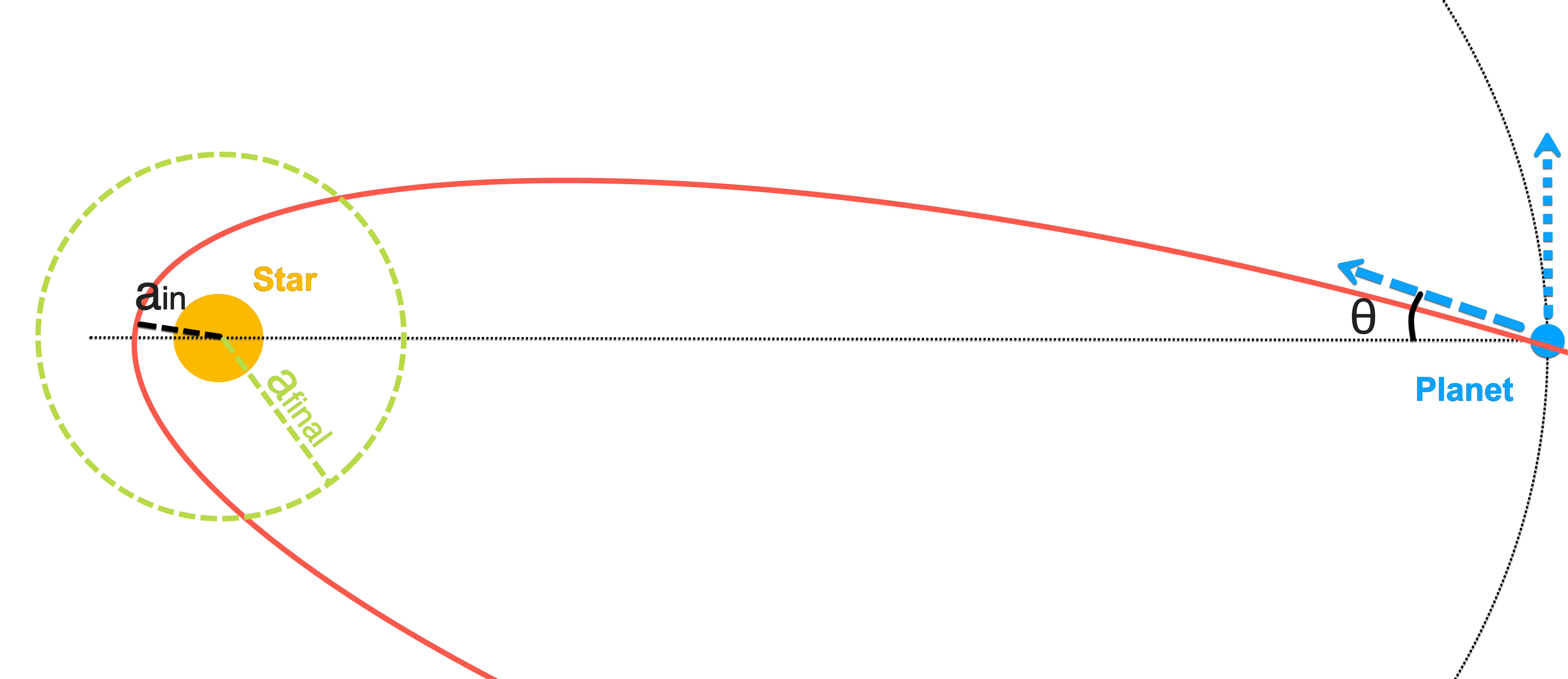}
\caption{This illustration shows a planet on an initial circular orbit (in black) is scattered inward towards to its host star, due to \emph{close encounter} and orbital angular momentum exchange with another massive object in the orbital plane, and transfers to a highly-eccentric orbit (in red). Later on, this \emph{eccentricity} of this highly-eccentric orbit gradually dampens down due to strong tidal interactions with its host star near its \emph{periastron}, and eventually the orbit circularizes into a final circular orbit (in dashed green) with semi-major axis ($a_{\text{final}}$).}
\label{fig:f3}
\end{figure}

We assume that before scattering the planet is in a circular orbit around its host star at a relatively large orbital separation $a(0)$, and that the scattering only changes the direction of its orbital velocity at the instant of the scattering, but does not change the magnitude of its orbital velocity. Thus, the original velocity vector lies in the orbital plane and is perpendicular to the direction of its host star. Immediately after the scattering event, the new velocity vector makes an angle $\theta$ with respect to the direction to its host star (Fig.~\ref{fig:f3}). If the planet is sufficiently close to the host star when it passes near the periastron ($a_{\text{in}}$) of this new eccentric orbit, the tides on the planet, which are raised by its host star, would be strong enough to dissipate the orbital energy and gradually circularize the orbit with semi-major axis $a_{\text{final}}$. This process of \emph{eccentricity dampening} due to tides approximately conserves the angular momentum of the orbit ($L_{\text{orbit}}$).



The \emph{eccentricity} $e$ of the eccentric (transitional) orbit is simply: 
\begin{equation}\label{eq9}
    e = \cos \theta = 1 - \frac{\theta^2}{2} + \frac{\theta^4}{24} +...,
\end{equation}

Thus, if $\bf{\theta}$ is small enough, then both $a_{\text{in}}$ and $a_{\text{final}}$ are proportional to $\theta^2$ because: 
\begin{eqnarray}\label{eq10}
    a_{\text{in}} &=& a \cdot (1-e) = a \cdot (1-\cos \theta) \approx a \cdot \frac{\theta ^2}{2} \propto \theta ^2 \\
    a_{\text{final}} &=& a \cdot (1-e^2) = a \cdot (1-\cos^2 \theta) \approx a \cdot \theta^2 \approx 2 \times a_{\text{in}} \nonumber\\
\end{eqnarray}

We assume that the scattering occurs within the disk plane. Furthermore, we assume that the scattering angle $\theta$ is random and its PDF probability distribution \textbf{PDF}$(\theta)$ is smooth. Strictly speaking, \textbf{PDF}$(\theta)$ needs not to be a constant, but we only require it to vary slowly in between 90 $+$ a few degrees and 90 $-$ a few degrees for large-angle (close-to-right-angle) scattering which sends the planet towards the star. For simile, it is like pointing a shot-gun towards a distant target. 

Then, the probability of scattering in the interval of $\theta$ to $\theta + d\theta$ is proportional to $d\theta$. Thus, the number of planets $d\textbf{N}$ that end up between $\theta$ and $\theta + d\theta$ after a single scattering event is:

\begin{equation}\label{eq11}
    d\textbf{N} \propto d\theta \propto d(\sqrt{a_{\text{in}}}) \propto d(\sqrt{a_{\text{final}}})
\end{equation}

These proportionalities characterize the $\sqrt{a}$-dependence of \emph{Super-Neptunes}. 

In reality, the planet after such a scattering event may not remain exactly in the original orbital plane. They can also be ejected from the system, as evidenced by the many known free-floating planets~\citep{Mroz2024}. This offset will be magnified when it gets close to its host star, resulting in some of them having high obliquity in the end and, thus, mis-alignment between host stellar spin and planet orbital plane axis. 

The absolute occurrence rates themselves also give hints to the \emph{origins} of these \emph{Super-Neptunes}. Eq.~\ref{Eq:dN10b} and Eq.~\ref{Eq:dN11b} indicate that the \emph{cumulative number} of such planets \emph{per host star} inside a certain orbital separation $a_{\text{out}}$ can be estimated as:

\begin{equation}\label{Eq:dN10b1}
   \textbf{N}_{4-10 \text{R}_{\oplus}} \approx B_{4-10 \text{R}_{\oplus}} \times \big(\sqrt{a_{\text{out}}} \big) \approx 0.1 \times \bigg( \sqrt{ \frac{a_{\text{out}}}{\text{AU}} } \bigg) \quad, a<a_{\text{out}}
\end{equation}

and,

\begin{equation}\label{Eq:dN11b1}
   \textbf{N}_{>10 \text{R}_{\oplus}} \approx B_{>10 \text{R}_{\oplus}} \times \big(\sqrt{a_{\text{out}}} \big) \approx 0.05 \times \bigg(\sqrt{ \frac{a_{\text{out}}}{\text{AU}} } \bigg) \quad, a<a_{\text{out}}
\end{equation}

The normalization constant, \emph{B}, is explained for Eq.~\ref{Eq:dN3b}, above. These orders of magnitude estimates are based on the assumption that $a^{1/2}$-dependence is carried out at larger orbital distances.

Thus, $a_{\text{out}}$ must be at least on the order of \emph{a few} AU or \emph{a few tens} of AU, to increase the \emph{cumulative number} of such planets to the order of unity. Then, from this perspective, our solar system is not an outlier but agrees very well with the statistics derived here, since all four giant planets (with R$_{\text{p}}\gtrsim$ 4 R$_{\oplus}$), two Gas Giants (\emph{Jupiter} and \emph{Saturn}), and two Ice Giants (\emph{Uranus} and \emph{Neptune}), lie at \emph{a few} AU or \emph{a few tens} of AU. Then, the location where these \emph{Super-Neptunes} originate is very likely beyond the \emph{snow-line}. This suggests that cosmic ices (H$_2$O, NH$_3$, CH$_4$, CO, etc.) likely play a major role in their formation and contribute significantly or even predominately to their core masses~\citep{PNAS:Zeng2019}, which is consistent with some recent JWST transmission spectral measurements~\citep{Barat2026}.

These \emph{Super-Neptunes} are \emph{far less populous} compared to smaller planets, but since they are more readily detectable by current observational methods (both transits and RV), they do show up distinctly in the current \emph{Mass-Radius plot}, \emph{Period-Radius plot}, and many other plots and statistics of exoplanets. 




A corollary of the scattering scenario is that it predicts a (strong) positive correlation between the host stellar metallicity [Fe/H] and the occurrence rate of close-in \emph{Super-Neptunes}. This positive metallicity \textbf{[Fe/H]} correlation is in fact confirmed by observation, e.g.,~\citep{Buchhave2014, Wang2015RevealingStars, Fischer2005TheCorrelation}, and explained in detail in the next section.  

\subsection{Correlation of Close-in Super-Neptune Occurrence Rate with Host-stellar Metallicity}

For simplicity of discussion, let us define the lower-case letter $n$ as the number of planet cores that are formed in a mutually interactive region around and beyond the snowline and are massive enough to scatter each other. Now, let us assume that this $n$ is proportional to the host stellar metal abundance (since the metallicity \textbf{[Fe/H]} is a logarithmic-base-10) as: 

\begin{equation}\label{eq13}
    n \propto 10^{\textbf{[Fe/H]}}
\end{equation}

Here we use \textbf{[Fe/H]} instead of \textbf{[M/H]} for the convenience of calculation because \textbf{[Fe/H]} is readily measured. If there is only one core, then there is no scattering. Therefore, at least two or more cores are required for the scattering to possibly occur:

\begin{equation}\label{eq14}
    \text{Probability of Scattering} \propto n \cdot (n-1)
\end{equation}

Let us assume that without changing other conditions, the cumulative number \textbf{N} of close-in \emph{Super-Neptunes} \emph{per host star} is also proportional to this Probability of Scattering. 

So, when $n$ is large enough, one expects to find: 
\begin{equation}\label{eq15}
    N \propto n \cdot (n-1) \sim n^2 \propto 10^{(2 \times \textbf{[Fe/H]})},
\end{equation}

Eq.~\ref{eq15} is a power-law dependence with power-index $\beta$ of 2 in log-base-10. Therefore,  

\begin{eqnarray}\label{eq16}
    \lg N &=& \beta \cdot \textbf{[Fe/H]} + \text{const}, \\
    \beta &\sim& +2 \nonumber
\end{eqnarray}

Generally speaking, for close-in \emph{Super-Neptunes} resulting from \emph{planet-planet scattering}: 

\begin{equation}\label{eq17}
    \beta \gtrsim +2
\end{equation}

because when $n$ is small, it renders an \emph{even stronger} dependence of the planet occurrence rate on \textbf{[Fe/H]}. Since going from one core to two cores, the \emph{Probability of Scattering} suddenly jumps from zero to a finite number. Thus, $\beta$ should be even higher for \emph{Sub-Saturns} than \emph{Jupiters}.

This is exactly what the authors of~\cite{Petigura2018ThePlanets} have found from their statistics:
\begin{itemize}
    \item \emph{Sub-Saturns}: $\beta \approx +5.5$
    \item \emph{Jupiters}: $\beta \approx +3.4$
\end{itemize}

Of course, we need to be aware of other physical mechanisms which could also give rise to the positive correlation between the occurrence rate of close-in \emph{Super-Neptunes} and their host-stellar metallicities. So, our analysis above is not exclusive to other possibilities. 

\subsection{Hot Jupiter Inflation}
Finally, the physical mechanisms of the radius inflation of Hot Jupiters can be explored, using their radius versus equilibrium temperature relations (Fig.~\ref{fig:f4}) and the correlation with the metallicity. 

In order to do that, let us plot the \emph{Kepler} planets in their Equilibrium Temperatures (T$_{\text{eq}}$) Versus Planet Radii (R$_{\text{p}}$), and among them, we color-code the \emph{Hot Jupiters} according to their host-stellar metallicities (\textbf{[Fe/H]}). See Fig.~\ref{fig:f4}. It shows that: 

\begin{figure*}
\centering
\includegraphics[width=\textwidth]{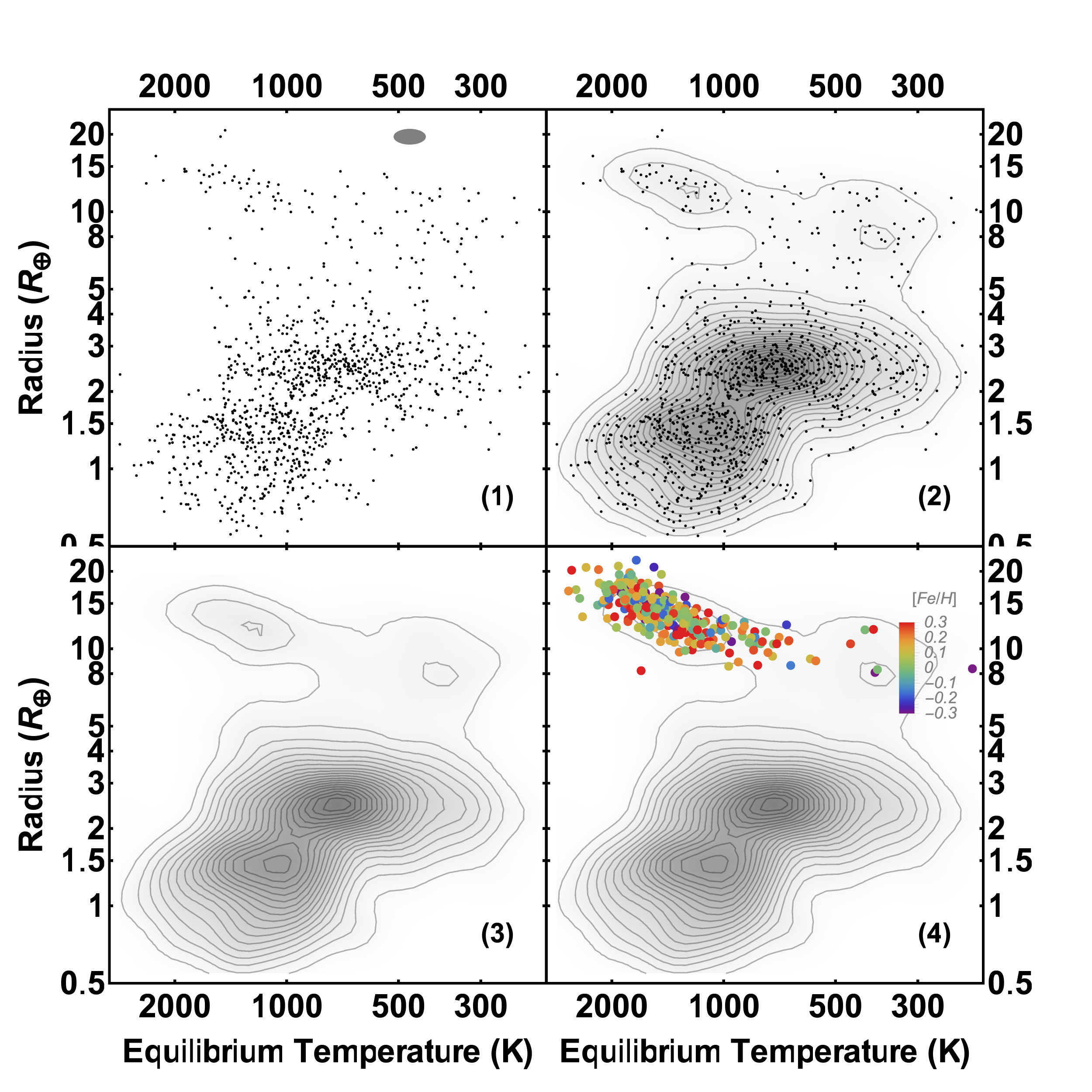}
\caption{Plotted here are the \emph{Kepler} confirmed planets/candidates with \emph{Gaia DR2} radius improvements. Selection criteria are main-sequence host-star with temperature 5000-6500 K and planet radius errors less than $\pm 10\%$. Data source: exoplanet archive~\protect\citep{Akeson2013TheResearch}~\protect\citep{Christiansen2025}~\protect\citep{Thompson2017Planetary25} and Gaia DR2~\protect\cite{Berger2020,Berger2018Revised2,GaiaCollaboration2018GaiaProperties,Lindegren2018}. Sub-Panels: (1): 1100$+$ planets fulfilling the criteria, with gray ellipse showing the typical errors in planet radius and temperature. (2) and (3): Smooth Kernel fit to data. (4): All known transiting \emph{Hot Jupiters} (R$_{\text{p}}>8$ R$_{\oplus}$) which have masses measured from RV surveys (e.g. WASP, HAT, HAT-P, etc.), colored by metallicity (\textbf{[Fe/H]}). also c.f.~\protect\citep{Castro-Gonzalez2024}.}
\label{fig:f4}
\end{figure*}

\begin{itemize}
\item First, the planet population most susceptible to host stellar irradiation is the \emph{Super-Neptune Sub-Saturn} population in between 4 and 8 R$_{\oplus}$, known as the \emph{Hot Neptune Dessert}. They are most likely the \emph{true} gas dwarfs--planets with light \textbf{Hydrogen/Helium}-dominated gaseous envelopes that are significantly less massive compared to envelopes of gas giants such as \emph{Saturn} or \emph{Jupiter}, and thus have relatively low surface escape velocity and are more susceptible to lose the gaseous envelopes~\citep{PNAS:Zeng2019}.
\item Second, this \emph{Super-Neptune Sub-Saturn} population seems to peak at much greater orbital distance from their host stars, which is inferred by a small but statistically significant peak at around $\sim 300$ Kelvin or even colder T$_{\text{eq}}$. This peak suggests that this population originated from greater orbital distances, likely near or beyond snowline, which is consistent with our previous analysis. They seem to have connections with both the \emph{Super-Earth Sub-Neptune} population and the \emph{Hot Jupiter} population, and form a \emph{bridge} between the two.
\item Third, the \emph{radius inflation} of \emph{Hot Jupiter}'s radius is nicely and positively-correlated with the Equilibrium Temperatures (T$_{\text{eq}}$) of the planets, and thus, the intensity of the host stellar bolometric radiation intensity incident onto the planet surface. This trend of \emph{radius inflation} towards higher T$_{\text{eq}}$ matches exactly for both the \emph{Kepler}'s \emph{Hot Jupiter} population and the \emph{RV}'s \emph{Hot Jupiter} population probed by a totally different method --- the ground-based radial-velocity surveys. Thus, the \emph{radius inflation} of \emph{Hot Jupiters} is directly caused by host stellar radiation energy dumped into the gaseous envelopes of these \emph{Hot Jupiters}. Recent observations even suggest that this \emph{radius inflation} is instantaneous compared to the age of these planetary systems, 
such as the recent observations of inflated planets around newly-evolved host stars~\citep{Lopez2016Re-inflatedGiants, Grunblatt2017SeeingStars}). The host stellar metallicity (\textbf{[Fe/H]}), on the other hand, has less of an effect on \emph{radius inflation}. 
\end{itemize}

Last but not least, if the hot Jupiter population inflates systematically with increasing host-stellar radiation, and if the main component of these sub-Neptunes' envelopes was dominantly gaseous Hydrogen or Helium, then the sub-Neptune population would trend more with the host-stellar radiation in Fig.~\ref{fig:f4}. However, we do not see this trend for the centrum of the sub-Neptune population. This is another piece of evidence supporting that the \emph{backbone} of these sub-Neptunes is the Oxygen-element---the third most abundant element in the current universe after Hydrogen and Helium. Most likely, this Oxygen exists chemically as H$_{2}$O in these sub-Neptunes. Thus, these water-rich sub-Neptunes and rocky super-Earths can be collectively called an Oxygen-planet or an Oxygen-dominated planet, a term suggested by V.M. Goldschmidt more than 100 years ago~\citep{Goldschmidt1954}. 

\section{Conclusion}
In this paper, we have elucidated mathematically the orbital distributions of the exoplanet population observed from the \emph{Kepler}'s primary mission, expressed as either the orbital period \emph{P} or the orbital semi-major axis \emph{a} space, \emph{per host star}, in a collective way, and also in various radius bins. Based on the method of \emph{survival function analysis}, we are able to derive analytic functional dependence of these orbital distributions, compare the results, and confirm the results with similar works in the field. We then visited several possible formation scenarios and pathways for planets in different size ranges, which can explain the results from a theoretical point-of-view. We conclude that 

\emph{The semi-major-axis distribution of super-Earths and sub-Nepunes is logarithmically flat, likely due to in-situ formation or disk migration. Meanwhile, the $\sqrt{a}$-flat distribution of super-Nepunes is likely due to inward scattering.}


\section*{Acknowledgement}
Li Zeng thanks Laura Kreidberg, Lars A. Buchhave, and Jinglin Zhao for discussion of the overall exoplanet distribution. This project has received funding from the Research Council of Norway through the Centres of Excellence funding scheme, project number 332523 (PHAB) and project number 360579 (WaterWorlds). Stein B. Jacobsen acknowledges support from the DOE-NNSA grant DE-NA0004231 to Harvard University: From Z to Planets: Phase V. 


 
\section*{Data availability}
The data underlying this article are available in \emph{NASA Exoplanet Archive} (~\url{https://exoplanetarchive.ipac.caltech.edu/}). The \emph{Wolfram Mathematica} codes used to process the data are or will become available in Wolfram Community Post, for example, under (~\url{https://community.wolfram.com/groups/-/m/t/3196285}) and (~\url{https://community.wolfram.com/groups/-/m/t/2445247}). The data and codes underlying this article will also be shared on reasonable request to the corresponding author. The authors reserve the rights to modify and improve the codes and data, and to incorporate new data as they become available. 

\bibliographystyle{mnras}
\bibliography{mendeley_v7}

\end{document}